\documentclass[10pt]{article}
\usepackage{amssymb,euscript,bbold}
\usepackage{amsfonts}
\usepackage{amssymb}
\usepackage{mathrsfs}
\usepackage{graphicx}
\usepackage{slashed}
\usepackage{amsthm}
\usepackage{amsmath}
\usepackage{amscd}
\usepackage{xcolor}
\usepackage{mathtools}

\DeclarePairedDelimiterX\braket[2]{\langle}{\rangle}{#1 \delimsize\vert #2}

\usepackage{float}
\usepackage[caption = false]{subfig}

\usepackage{geometry}
 \usepackage[parfill]{parskip}

\newcommand{\be}{\begin{equation}}
\newcommand{\ee}{\end{equation}}
\newcommand{\ba}{\begin{eqnarray}}
\newcommand{\ea}{\end{eqnarray}}

\def\P{\mathbb{P}}
\def\Z{\mathbb{Z}}
\def\R{\mathbb{R}}
\def\C{\mathbb{C}}

\theoremstyle{definition}

\newtheorem{example}[subsubsection]{Example}

\theoremstyle{remark}

\newtheorem*{remark*}{Remark}

\newcommand{\Rr}{\mathbb{R}} 
\newcommand{\Pp}{\mathbb{P}}

\usepackage{hyperref}
\hypersetup{hidelinks,backref=true,pagebackref=true,hyperindex=true,colorlinks=false,breaklinks=true,urlcolor= ocre,bookmarks=true,bookmarksopen=false,pdftitle={Title},pdfauthor={Author}}

\begin{document}
\input{epsf}

\begin{flushright}
KCL-PH-TH/2020-64
\end{flushright}

\begin{flushright}

\end{flushright}
\begin{flushright}
\end{flushright}
\begin{center}
\Large{ New $G_2$-conifolds in $M$-theory and their Field Theory Interpretation}\\
\bigskip
\large{B.S. Acharya*, L. Foscolo**, M. Najjar* and E.E. Svanes***}\\
\smallskip\normalsize*{\it
Abdus Salam International Centre for Theoretical Physics, Strada Costiera 11, 34151, Trieste, Italy and Department of Physics, Kings College London, London, WC2R 2LS, UK}\\
\smallskip**{\it
Department of Mathematics, University College London, London, WC1E 6BT, UK}\\
\smallskip***{\it
Department of Mathematics and Physics, University of Stavanger, Kristine Bonnevies vei 22, 4021 Stavanger, Norway}\\
\end{center}

\bigskip
\begin{center}
{\bf {\sc Abstract}}
\end{center}

A recent theorem of Foscolo-Haskins-Nordstr\"om \cite{foscolo2017complete} which constructs complete $G_2$-holonomy orbifolds from circle bundles over Calabi-Yau cones can be utilised to construct and investigate a large class of generalisations of the $M$-theory flop transition. We see that in many cases a UV perturbative gauge theory appears to have an infrared dual described by a smooth $G_2$-holonomy background in $M$-theory. Various physical checks of this proposal are carried out affirmatively.

\newpage

\section{Introduction}
${\cal N}=1$ supersymmetric Minkowski $M$-theory vacua in which the seven extra dimensions of space are modelled on a space with $G_2$-holonomy provide a broad framework for addressing particle physics \cite{Acharya:2001gy,Friedmann:2002ty,Acharya:2005ez}, supersymmetry breaking \cite{Acharya:2008zi}, dark matter \cite{Acharya2018} and cosmology \cite{Kane:2019nod}. In $M$-theory on a $G_2$ holonomy space $X$, the basic ingredients of particle physics in four dimensions, namely non-Abelian gauge fields and chiral fermions arise from codimension four and codimension seven singularities respectively \cite{Acharya:2001gy, Acharya:2000gb}. 

The requisite singularities and the fact that the problem of constructing $G_2$-holonomy spaces is difficult even without singularities has previously meant that such vacua of $M$-theory are less well understood than those based on Calabi-Yau manifolds. Despite this, in recent years, there has been significant progress in understanding aspects of both the physics and mathematics of $G_2$-holonomy spaces, including new constructions of spaces with $G_2$-holonomy.

Advances have been made both from the physics \cite{Braun:2016igl, Guio:2017zfn, Braun:2017uku, Braun:2018joh, Heckman:2018mxl, delaOssa:2019jsx} and the mathematics \cite{foscolo2015new, foscolo2017complete, foscolo2018infinitely, joyce2017new, cheng2019bubble, dwivedi2019gradient, ball2019mean, acharya2019circle}, and involves aspects such as a string-worldsheet interpretation \cite{Fiset:2017auc, delaOssa:2018azc, Fiset:2018huv}, a better understanding of exceptional gauge bundles and instantons \cite{walpuski2012g_2, sa2015g2, delaOssa:2016ivz, delaOssa:2017pqy, haydys2017g2, joyce2018canonical, paladines2019instantons}, in addition to non-perturbative aspects such as $M2$ instantons and their relation to exceptional enumerative geometry \cite{Joyce:2016fij, Braun:2018fdp, Acharya:2018nbo}. On the physics side such new insights often rely on dualities with other string theories and F-theory \cite{Braun:2017ryx, Braun:2017csz, Braun:2019lnn, Braun:2019wnj}, in addition to dimensional reductions \cite{Braun:2018vhk, Barbosa:2019bgh, Barbosa:2019hts, Hubner:2020yde, Acharya:2020xgn}. \\

In this paper we will apply a recent and powerful new construction of non-compact $G_2$-holonomy manifolds and orbifolds \cite{foscolo2017complete} to $M$-theory. These examples are motivated from $M$-theory/Type IIA duality because they arise as circle bundles over asymptotically conical Calabi-Yau threefolds. In fact, they are constructed perturbatively in the small radius limit of the $M$-theory circle or equivalently at weak string coupling. The main theorem of \cite{foscolo2017complete} requires a topological condition on the circle bundle to be satisfied and we interpret this physically as requiring a critical point of the flux superpotential in Type IIA theory. Our main focus will be on an infinite set of examples which can be thought of as new conifolds in $M$-theory. They are all partial or complete desingularisations of an infinite set of conically singular $G_2$-holonomy spaces. In fact, as we will show, a fixed $G_2$-holonomy cone from this set can have a {\it variety of topologically distinct desingularisations}, a fact which has a natural physical interpretation: they correspond to distinct classical vacua of a perturbative UV gauge theory; simultaneously we show that there is a corresponding IR picture of these vacua in terms of a smooth $G_2$-holonomy space which has the interpretation in Type IIA string theory as different flux configurations on a fixed Calabi-Yau threefold. We check this latter interpretation by matching the number of unbroken $U(1)$ factors in the gauge group both in the UV and IR. The UV and IR descriptions correspond to, respectively, the open and closed string descriptions in Type IIA string theory. 

All of the examples we consider are natural generalisations of the original examples of the $M$-theory flop considered in \cite{Acharya:2000gb, Atiyah:2000zz, Atiyah:2001qf} so we will begin by reviewing relevant aspects of that discussion for the ease of the reader. Following that we describe the Calabi-Yau threefolds in Type IIA that we will consider and  the classical superpotential. In section three we describe the new $G_2$-holonomy spaces in detail and their connection to UV gauge theories on the open string side. We discuss the topological requirement which arises from the superpotential and describe some examples. In section four we describe the open/closed or UV/IR correspondences that arise in detail by matching the vacua of the UV theory to particular smooth $G_2$-manifolds in the IR. A number of examples are studied for illustrative purposes.

\subsection{Models for codimension four singularities and the $M$-theory flop}
We begin with a short review of the original $M$-theory flop and its reduction to type IIA string theory \cite{Acharya:2000gb, Atiyah:2000zz, Atiyah:2001qf}. Our main results are a significant generalisation of this.
Elegant models for the behaviour of $X$ near a codimension four orbifold singularity exist which we will recall here. Bryant and Salamon constructed complete $G_2$-holonomy metrics on the total space of $S( S^3)$, the spin bundle of the 3-sphere \cite{Bryant:Salamon}. These metrics are asymptotically conical and approach a metric cone $C( S^3 \times  S^3) \equiv \R^+ \times  S^3 \times  S^3$ over the unit spinor bundle, where the metric on $ S^3 \times  S^3$ is the homogeneous metric with respect to ${SU(2) \times SU(2) \times SU(2) \over \Delta SU(2)}$, with $\Delta SU(2)$ the diagonal $SU(2)$ subgroup.
One can view the cone, $C( S^3 \times  S^3)$, as $S( S^3)$ minus the zero-section or as a limit of $S( S^3)$ when the zero-section is collapsed to a point. In fact, in the Bryant-Salamon metric, the zero-section is an associative submanifold and its size is a parameter of the metric. 
A key point is that, in desingularising $C( S^3 \times  S^3)$ to the smooth $S( S^3)$ {\it there is a discrete choice of which} $ S^3 \equiv SU(2) \subset SU(2)^3/SU(2)$ one picks to complete the space, so the metrics at infinity have an additional threefold permutation symmetry, $S_3$, broken to $\Z_2$ by picking a sphere and desingularising. The moduli space of $G_2$-holonomy metrics is thus represented as three positive real lines emanating from the conical singularity, representing the three ways in which the conical singularity can be resolved.\\ 

\begin{figure}[H]
\begin{center}
\subfloat[The quantum moduli space.]{\includegraphics[width = 2.2in]{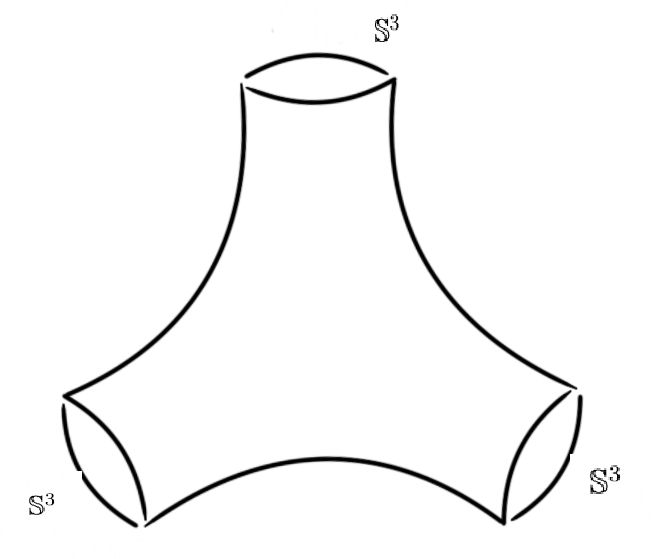}}
\label{fig:modulispace}

\subfloat[$M$-theory on an ADE-quotient of S($\mathbb{S}^{3}$)]{\includegraphics[width = 2.6in]{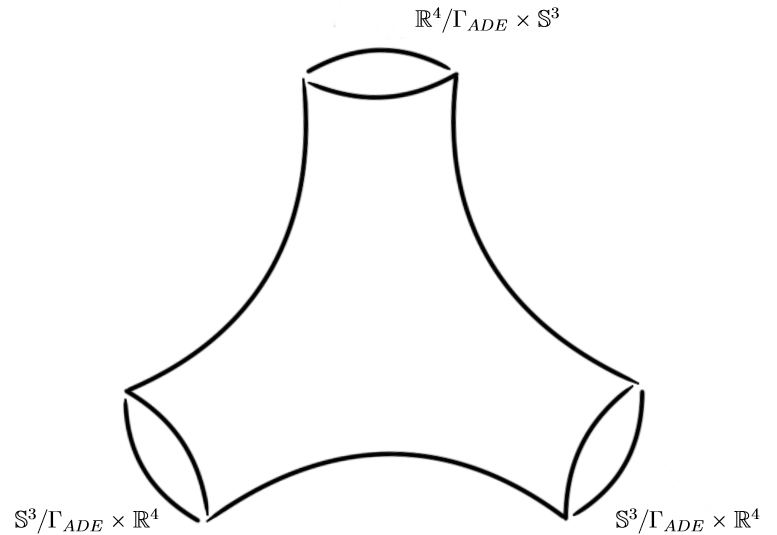}} \;\;\;\;
\subfloat[The type-IIA description of figure(d).]{\includegraphics[width = 3.5in]{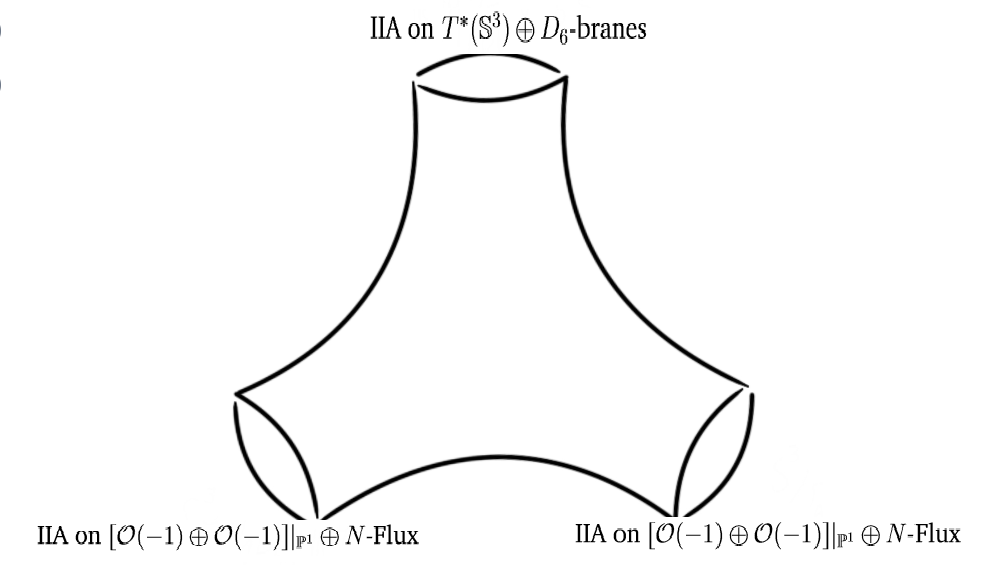}}
\label{fig:quotientmodulispace}
\end{center}
\caption{The story of moduli space of $M$-theory on the Bryant–Salamon $G_{2}$-space. }
\label{modulispace$M$-theory}
\end{figure}

In $M$-theory this space is enhanced to a complex one-dimensional space by the $M$-theory three-form field $C\in\Omega^3(X)$, with complex parameter
\begin{equation}
z={\rm Vol}( S^3)\,e^{i\int_{ S^3}C}\:,
\end{equation}
where
\begin{equation}
{\rm Vol}( S^3)=\int_{ S^3}\varphi\:,
\end{equation}
and $\varphi$ is the positive three-form defining the $G_2$ structure. Viewing this as a (quantum) moduli space of theories with three classical geometric branches, Atiyah and Witten further argued that the singular moduli space enhances to a smooth one \cite{Atiyah:2001qf}. The three classical branches of the moduli space are permuted by the $M$-theory flop \cite{Atiyah:2000zz}. This is depicted in Figure \ref{fig:modulispace} - a, b, c.\\

To introduce codimension four orbifold singularities and, hence, Yang-Mills fields, one can consider quotients of this space by discrete subgroups of the three $SU(2)$ symmetries $\Gamma_{ADE} \subset SU(2)^3$, which acts differently on the three branches of moduli space. 
One of the classical branches takes the form $\R^4/\Gamma_{ADE}\times  S^3$. This corresponds to an effective ultra violet (UV) gauge theory of $ADE$ type on $ S^3\times \R^{(3,1)}$ which reduces to pure $N=1$ super Yang-Mills theory at low enough energies. The other two branches of the moduli space take the form $\R^4\times  S^3/\Gamma_{ADE}$, and may be viewed as two infra red (IR) phases of the previously mentioned UV theory. Since the quantum moduli space is expected to be smooth and connected, this allows one to demonstrate confinement and the existence of domain walls and a mass gap, since the latter two spaces are smooth $G_2$-manifold and amenable to a semi-classical analysis. We will see many more examples of this behaviour in this paper, namely we will consider singular $G_2$-holonomy spaces which support non-Abelian 4d ${\cal N}=1$ supersymmetric gauge theories and show that there are topological transitions of these spaces to smooth $G_2$-holonomy manifolds. We will be able to compute some physical quantities on both sides of the transition, providing evidence that the transitions occur on a connected, smooth component of the quantum moduli space.\\

There is a Type IIA limit of the above $M$-theory flop transitions that will also play a role in our new results. These played a role in \cite{Acharya:2000gb,Atiyah:2000zz, Atiyah:2001qf} but strictly speaking at that time there was no known $G_2$-holonomy metric corresponding to the weakly coupled Type IIA limit as the Bryant-Salamon metrics are asymptotically conical and, hence, the $M$-theory circle has infinite radius at infinity. However, more general $G_2$-holonomy metrics on $S( S^3)$ with $U(1) \times SU(2)^2$ symmetry \cite{Brandhuber:2001yi} and finite circle at infinity were constructed later.
In $\R^4/\Z_N\times  S^3$ if we choose the $M$-theory circle to be in the $\R^4$ factor (viewed as a circle bundle over $\R^3$ minus the origin), the quotient is actually the deformed conifold $T^* S^3$ with $N$ $D6$ branes wrapping the $ S^3$. This will thus sometimes be referred to as the open string side. On the other hand, in the two flops of the form $\R^4\times  S^3/\Z_N$ the $M$-theory circle is the Hopf fibre of the $ S^3$ and hence the IIA limit is $\R^4\times\P^1$, which we can view as the resolved conifold with $N$ units of Ramond-Ramond flux on the two-sphere. This is thus referred to as the closed string side since there are no $D$-branes. These two IR phases are geometrically distinguished by the usual flop of the resolved conifold, while the UV/IR transition corresponds to the geometric conifold transition. This is depicted in Figure \ref{fig:quotientmodulispace} - d, e.\\

One may ask about more general discrete quotients of these $G_2$-holonomy backgrounds and their flops.
In particular, we will ask what happens when in type IIA we consider a further quotient of the $\R^4$ fiber of the resolved conifold by another $\Gamma_{ADE}$. For $\Gamma_{ADE}=\Z_p$ these have been dubbed hyperconifolds \cite{Davies:2009ub, Davies:2011is, Davies:2013pna}, and we will mostly restrict to such quotients in the current paper. We will consider the corresponding effective theories, ask about the lift to $M$-theory and the corresponding $G_2$-holonomy metrics and investigate the $M$-theory flop in this setting and the corresponding UV dynamics which turns out to be governed by $SU(N)$ theories on Lens spaces $ S^3/\Z_p$. 

\begin{figure}[H]
\begin{center}
\subfloat[The geometric classical moduli space.]{\includegraphics[width = 2.2in]{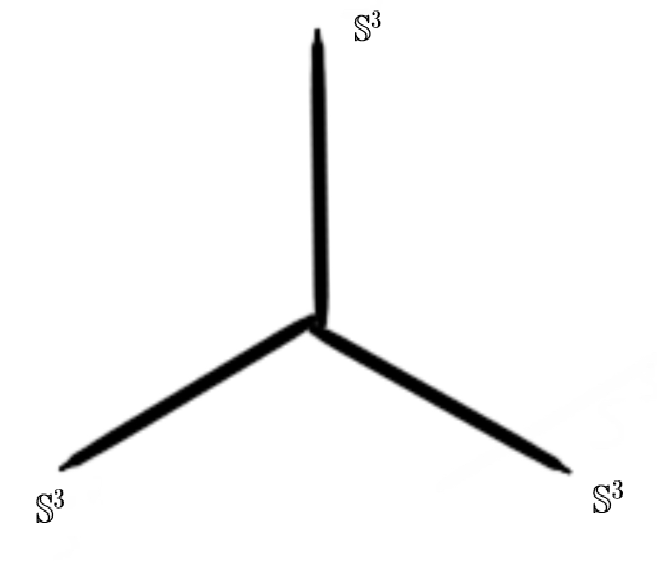}} \;\;\;\;
\subfloat[The classical moduli space due to $M$-theory.]{\includegraphics[width = 2.2in]{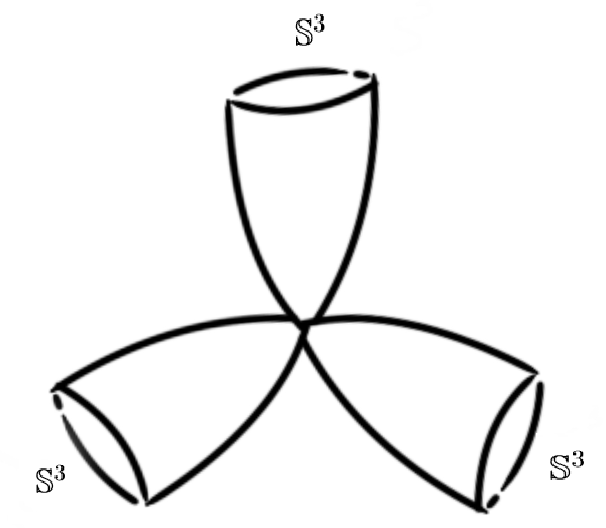}} \;\;\;\;
\end{center}
\end{figure}

\section{Type IIA Theory on Resolved Hyperconifolds}

The conifold in three complex dimensions is defined as the following hypersurface in $\C^4$:
\be
{\cal C} \equiv \{(z_1, z_2, z_3, z_4) \in \C^4 | \sum_{i=1}^4 z_i^2 = 0 \}
\ee

${\cal C}$ is a cone whose cross-section is $S^2 \times  S^3$, ${\cal C} = C(S^2 \times  S^3)$. The singular apex of the cone can be {\it resolved} by replacing it with a holomorphic $\P^1$. The resolved conifold $\widetilde{\cal C}$ is the total space of a complex vector bundle over $\P^1$:
\be
\widetilde{\cal C} \equiv {\cal O}(-1) \oplus {\cal O}(-1)|_{\P^1}
\ee

A complete metric of $SU(3)$ holonomy on $\widetilde{\cal C}$ was explicitly constructed by Candelas and de la Ossa \cite{Candelas:1989js}. This metric is asymptotically conical. $\widetilde{\cal C}$ is a natural model for the normal bundle of a rigid holomorphic $\P^1$ in a generic smooth Calabi-Yau threefold. One can think of ${\cal C}$ as $\widetilde{\cal C}$ minus the zero section. \\

The Candelas-de la Ossa metric has two $SU(2)$ symmetries and, therefore, one can consider  both $\widetilde{\cal C}/\Gamma$ and ${\cal C}/\Gamma$, where $\Gamma \subset SU(2)^2$ is a finite subgroup of the isometry group. The singular Calabi-Yau threefolds ${\cal C}/\Gamma$ were dubbed {\it hyperconifolds} by \cite{Davies:2013pna} and \cite{Anderson:2018heq}. We will mostly be interested in the following hyperconifolds:
\be
\widetilde{\cal C}/\Gamma_{ADE} \equiv {{\cal O}(-1) \oplus {\cal O}(-1) \over \Gamma_{ADE}}|_{\P^1}
\ee
where $\Gamma_{ADE}$ is a finite subgroup of $SU(2)$ acting purely on the $\C\oplus\C\equiv\C^2$ fibres of $\widetilde{\cal C}$. Hence the notation on the right hand side.
${{\cal O}(-1) \oplus {\cal O}(-1) \over \Gamma_{ADE}}|_{\P^1}$ has an $ADE$ singularity supported on the zero section $\{0\} \times \P^1$. In Type IIA theory on 
$\widetilde{\cal C}/\Gamma_{ADE}$, at energies low compared to the inverse size of the zero section, the physics is described by ${\cal N}=2$ super Yang-Mills theory with $ADE$ gauge group and no hypermultiplets. The complexified gauge coupling is given by
\be
\tau = {4\pi \over g^2} + 2\pi i \theta = 2\pi \int_{\P^1} \omega + i B
\ee
where $\omega$ is the K{a}hler form of the AC Calabi-Yau metric and $B$ the Type IIA $B$-field.  $\widetilde{\cal C}/\Gamma_{ADE}$ is the ${\cal N}=2$ analogue of ${S( S^3)/ \Gamma_{ADE}}$ in $M$-theory.\\ 

In the correspondence between Type IIA theory on $\widetilde{\cal C}/\Gamma_{ADE}$ and ${\cal N}=2$ $ADE$ super Yang-Mills theory, the classical moduli space of vacua of the gauge theory may be identified, for fixed $\tau$, with the complexified K\"ahler moduli space of $\widetilde{\cal C}/\Gamma_{ADE}$. Perturbative and non-perturbative instanton corrections give rise to a quantum moduli space \cite{Seiberg:1994rs}. The gauge theory instantons may be identified with world-sheet instantons wrapping the zero section of $\widetilde{\cal C}/\Gamma_{ADE}$ and the quantum moduli space is perhaps most naturally incorporated by mirror symmetry as the complex structure moduli space of the mirror of 
$\widetilde{\cal C}/\Gamma_{ADE}$. This string theory picture of Seiberg-Witten theory was developed in \cite{Kachru:1995fv}.

\subsection{$M$-theory/Type IIA duality, ${\cal N}=1$ theories and New $G_2$-manifolds}

Type IIA theory on $\widetilde{\cal C}/\Gamma_{ADE}$ lifts to $M$-theory on $S^1 \times\widetilde{\cal C}/\Gamma_{ADE}$, with a metric on the 7-manifold which is a Riemannian product,
\be
ds_7^2 = R^2 dy^2 + g_6(\widetilde{\cal C}/\Gamma_{ADE})
\ee
with $R$ the size of the $M$-theory circle and $g_6(\widetilde{\cal C}/\Gamma_{ADE})$ the asymptotically conical (AC) Calabi-Yau metric on $\widetilde{\cal C}/\Gamma_{ADE}$. One may view this metric as a metric with holonomy contained in $G_2$. From this point of view, the Type IIA world-sheet instantons on $\{0\} \times \P^1$ lift to membrane instantons wrapping the associative 3-cycle $S^1 \times \{0\} \times \P^1$ at the orbifold point in the moduli space.\\

The $G_2$ point of view is extremely useful in contemplating a more general situation. Imagine that, up to higher order terms for small $R$, one can replace the metric above by

\be
ds_7^2 = R^2 \theta^2 + g_6(\widetilde{\cal C}/\Gamma_{ADE}) + \dots,
\ee

which is now a metric on a more general circle bundle over $\widetilde{\cal C}/\Gamma_{ADE}$ with connection $\theta$. So, in this case, $S^1 \rightarrow X_7 \rightarrow \widetilde{\cal C}/\Gamma_{ADE}$.
Following Apostolov-Salamon \cite{Apostolov_2004}, Foscolo-Haskins-Nordstr\"om (FHN) \cite{foscolo2017complete} studied holonomy $G_2$ metrics of this kind perturbatively in small $R$ and 
their main result is a powerful theorem which essentially states that, given a complete AC Calabi-Yau threefold, $Z$, and providing the circle bundle over $Z$ satisfies a certain topological condition, complete $G_2$-holonomy metrics exist on $X$. \\

There is a natural physical interpretation of this result. $M$-theory vacua which are isometric circle bundles can be interpreted in Type IIA as backgrounds with non-zero Ramond-Ramond 2-form flux, $F=d\theta$. $F$ is the curvature of the $M$-theory circle bundle. If, in addition, the $M$-theory background has holonomy $G_2$, then the IIA background may be interpreted at leading in order in $R$ as a Calabi-Yau threefold with $F$-flux. Switching on $F$ induces a potential for the K\"ahler moduli of $Z$, which is known to be induced from a superpotential \cite{Vafa:2000wi}
\be
W \propto \int_Z F\wedge T\wedge T\:.
\ee
Here, $T=B+i\,\omega$ is the complexified hermitian form, where $\omega$ is the K\"ahler form associated to the AC Calabi-Yau metric and $B$ is the internal part of the Kalb-Ramond two-form field. Supersymmetric vacua in the limit of decoupled gravity must satisfy $dW=0$ and this is equivalent to the condition that $F\wedge \omega = 0$ in $H^4(Z, \R)$ when the $B$-field background is switched off. In other words, $c_1(X) \cup \omega = 0$ in cohomology. This is precisely the topological condition in the Theorem of FHN. We see here that it has a very natural physical interpretation as a necessary condition for a supersymmetric (= $G_2$-holonomy) vacuum. Thus, starting with the $N=2$ theory defined by $Z$, switching on the RR flux $F$ induces a potential for the K\"ahler moduli fields whose critical points correspond to $G_2$-holonomy spaces in $M$-theory.

\section{New $G_2$-manifolds and SU(p) Gauge Theories}
In this section, we will study $G_2$-holonomy spaces constructed as circle bundles over $\widetilde{\cal C}/\Gamma_{A_{p-1}}$ and its resolutions $\widetilde{\widetilde{\cal C}/\Gamma_{A_{p-1}}}$ in more detail. The orbifold singularities of $\widetilde{\cal C}/\Gamma_{A_{p-1}}$ are along the zero section and are locally modelled on ${\C^2}/{\Z_p}$. In fact, $\widetilde{\cal C}/\Gamma_{A_{p-1}}$ is toric and can be resolved using standard techniques \cite{Xie:2017pfl}. Details are given below. Furthermore, these toric examples may be shown to admit complete AC Calabi-Yau metrics \cite{Futaki:Ono:Wang, VanCoevering:Existence,Goto:Crepant,Conlon:Hein:I} and hence can be used in the FHN theorem. In the simplest resolution, one replaces the zero section $\{0\}\times\P^1$ by a fibration over $\P^{1}$ with fibre $\Sigma_{p-1}$ a chain of $p-1$ $\P^1$s which intersect according to the Dynkin diagram of $SU(p)$. The sizes of these $\P^1$s physically correspond to the Coulomb branch moduli of ${\cal N}=2$ $SU(p)$ super Yang-Mills theory \footnote{Actually, in general there are further, topologically distinct resolutions for a given $p\geq3$. We will comment on these additional resolutions further below.}.\\

The would-be $G_2$-holonomy spaces are $S^1$ bundles over these resolutions. The circle bundle is defined by specifying $p$ integers, $\kappa_0,\kappa_1,\kappa_2,..,\kappa_{p-1}$, which are the quantised values of $F$-flux through each of the corresponding $\P^1$s. We can hence denote the 7-manifolds as $X_{(\kappa_0,\kappa_1,..,\kappa_{p-1})}$. Then, one has to examine the $c_1(X)\wedge \omega =0$ condition in detail as a function of the K\"ahler moduli. Some details of this are described below. Additionally, one has to specify boundary conditions on $F$ at infinity. The natural condition in Type IIA theory is that the RR gauge field is flat at infinity in $Z$. In our examples, $Z_{\infty} =  S^3/\Gamma_{ADE} \times \P^1$, which admits non-trivial flat connections. For $\Gamma_{A_{p-1}}=\Z_p$ this is equivalent to picking one of $p$ irreducible representations of $\Z_p$. If $\lambda = e^{2\pi i \over p}$, then a flat connection is a choice of $\lambda^q$, with $q \in \{0,..,p-1\}$. The total space of the circle bundle at infinity is then topologically $( S^3 \times  S^3)/\Gamma_{(p,N,q)}$, with $\Gamma_{(p,N,q)}$ a finite group depending on $p, N$ and a choice of flat connection $q$. Thus, all of the spaces $X_{(\kappa_0,\kappa_1,..,\kappa_{p-1})}$ are resolutions of the {\it same} fixed cone $\R^+ \times ( S^3 \times  S^3)/\Gamma_{(p,N,q)}$. If, in addition, the conditions in the FHN theorem are satisfied, we get {\it a host of topologically distinct 7-manifolds with $G_2$-holonomy metrics, resolving the same conical singularity.}\\

We will begin our discussion of the physics with one example, namely, 
$X_{(\kappa_0,0,\dots,0;\ q)}$.

\subsection{$M$-theory on $X_{(\kappa_0,0,..,0;\ q)}$}

As we will see below, the topological condition can be satisfied by setting all the sizes of the $\P^1$s in $\Sigma_{p-1}$ to zero. Hence, the space has a codimension four $A_{p-1}$ orbifold singularity. Globally, this means that $X_{(\kappa_0,0,..,0;0)} = {\R^4/\Z_p \times  S^3/{\Z_{\kappa_0}}}$ where the second factor is the circle bundle of degree $\kappa_0$ over the zero section $\P^1$. If we extend this to non-zero $q$, then the space becomes $X_{(\kappa_0,0,..,0;\ q)} = S( S^3)/\Gamma_{(p,N,q)}$ with $\kappa_{0} = N$. 
The group $\Gamma_{p,N,q}$ is described as
\begin{equation}
\Gamma_{p,N,q}= \{(\eta,\xi)\in\Z_p\times U(1)\;\vert\;\eta^q=\xi^N\}
\end{equation}
Here $\eta$ acts on the fibre of $\R^4$ and $\xi$ on the base $ S^3$ of $S( S^3)=\R^4\times S^3_2$. Note also that $\xi\in\Z_{rN}$, where $r$ is defined by
\begin{equation}
p=r\,\textrm{gcd}(p,q)
\end{equation}
To see why $\Gamma_{p,N,q}$ is as claimed, note that we can write
\begin{equation}
X_{(\kappa_0,..,0;\ q)}=\left(\R^4\times S^3_2\times S^1\right)\Big/\Z_p\times U(1)
\end{equation}
where $\zeta\in\Z_p$ and $\xi\in U(1)$ acts on the last factor as $\zeta^q\xi^{-N}$.

These examples are all quotients of the $B_7$ and $D_7$ metrics of \cite{Brandhuber:2001yi, Cvetic:2001ih} on $S( S^3)$. The case $p=2$ was studied in \cite{Hosomichi:2005ja}. Notice that the case $X_{(\kappa_0=1,0,\dots,0;\ 0)}={\R^4/\Z_p \times  S^3}$ and is essentially the case discussed in the introduction. Thus, we see that all of these examples are generalisations, essentially of the (quotients of) Bryant-Salamon examples.\\

Let's consider the $M$-theory flop of $X_{(\kappa_0,0,\dots, 0;\ 0)} = \R^4/\Z_p \times  S^3/{\Z_{\kappa_0}}$ to $\R^4/\Z_{\kappa_0} \times  S^3/{\Z_p}$ as described in section \ref{sec:OpneClosed}. The physics of $M$-theory on this space can be described as follows. The space has a codimension four orbifold singularity along $\{0\}\times  S^3/{\Z_{p}}$. Hence, there is a super Yang-Mills theory with gauge group $SU(\kappa_0)$ localised along this 3-manifold. This description is valid for large volume and energies below the inverse size of the the $ S^3$. Even though the energies are below the Kaluza-Klein scale, we will refer to this as the UV theory following the discussion in the introduction, since we will see that it will lead to a confining IR theory at much lower energies. To really specify the UV theory, one has to supply some additional data. In studying perturbative Yang-Mills theory on $ S^3/{\Z_{p}}$ one has to specify a choice of flat $SU(\kappa_0)$ connection on $ S^3/{\Z_{p}}$. From the lower energy point of view this is a choice of adjoint Higgs field. Such connections are classified by representations $R_i : {\Z_{p}} \rightarrow SU(\kappa_0)$ modulo gauge transformations. Essentially they can be transformed into diagonal $SU(\kappa_0)$ matrices of order $p$ up to permutations of the entries (Weyl group). The choice of such a flat connection breaks $SU(\kappa_0)$ to a product of $U(1)$’s and $SU$ factors. The latter is really the gauge group of the low energy, four-dimensional theory. Thus, since there are many flat connections, there are many UV theories. In the rest of the paper we investigate to what extent they all have IR duals which are smooth $G_2$-manifolds.\\

As explained above, $X_{(\kappa_0,\kappa_1,..,\kappa_{p-1};\ q)}$ is asymptotic to the {\it same} cone at infinity as $X_{(\kappa_0,0,..,0;\ q)} = S( S^3)/\Gamma_{(p,N,q)}$, namely $\R^+ \times ( S^3 \times  S^3)/\Gamma_{(p,N,q)}$. Hence all of these manifolds are resolutions of the {\it same} conical singularity, a necessary condition for the flop transition from the gauge theory to the IR phase to make sense.

As we will see, each choice of flat connection in the UV then corresponds to a choice of distribution of RR fluxes $X_{(\kappa_0,\kappa_1,..,\kappa_{p-1};\ q)}$ for the IR geometry. In addition, there is a distinct $SU(\kappa_0)$ gauge theory for each choice of $q\in\{0,1,..,p-1\}$. How exactly this matching between UV and IR geometries works is the subject of the remainder of the paper. For the IR geometries we use toric geometry to describe them as circle bundles over resolutions of the hyperconifold through the FHN theorem. Note however, as we will see later, that it is not always possible to solve the topological condition in the FHN theorem with strictly semi-positive values for the moduli fields i.e. a few solutions which exist formally are beyond the regime where the geometric approximation is valid.

\subsection{Toric Crepant Resolutions of $\widetilde{\cal C}/\Gamma_{A_{p-1}}$}

In order to analyse the conditions arising from the superpotential we will rely on the fact that the Calabi-Yau orbifolds we begin with are toric. The discussion will have to be rather technical in this subsection.
Starting with the conifold $\mathcal{C}$ \cite{Candelas:1989js}, we will consider some basic facts about its toric geometry. For a good introduction to toric geometry the reader is referred to e.g. \cite{fulton1993introduction}. Recall that as a hyper-surface in $\C^4$, the conifold can be written as
\begin{equation}
\label{eq:conifold}
y_1y_4-y_2y_3=0\:,
\end{equation}
where $(y_1,y_2,y_3,y_4)\in\C^4$. To see $\mathcal{C}$ as a toric variety, one makes the following substitution
\begin{equation}
y_1=z_1z_3\:,\;\;\;y_2=z_1z_4\:,\;\;\;y_3=z_2z_3\:,\;\;\;y_4=z_2z_4\:,
\end{equation}
which satisfies \eqref{eq:conifold} identically. For these new {\it homogeneous} coordinates we can implement the equivalence relation
\begin{equation}
(z_1,z_2,z_3,z_4)\sim(\eta z_1,\eta z_2,\eta^{-1}z_3,\eta^{-1}z_4)\:,\;\;\;\eta\in\C^*\:.
\end{equation}

\begin{figure}[H]
\begin{center}
\subfloat[The singular conifold.]{\includegraphics[width = 1.3in]{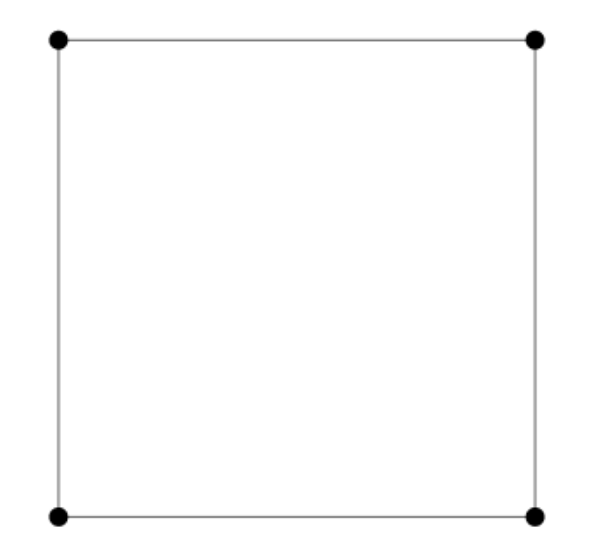}} \;\;\;\;
\subfloat[One toric resolution of the conifold.]{\includegraphics[width = 1.3in]{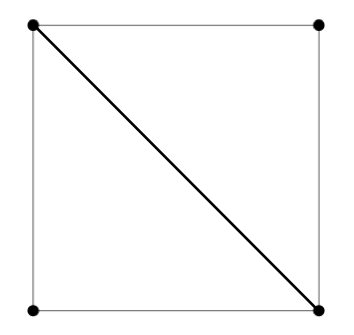}} \;\;\;\;
\subfloat[Another resolution of the conifold.]{\includegraphics[width = 1.3in]{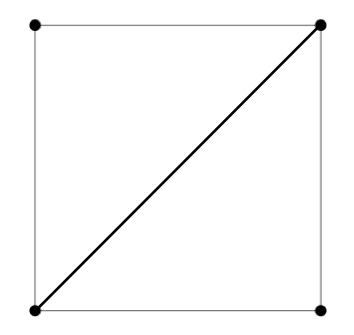}}
\end{center}
\caption{Toric diagrams of the conifolds and its two different resolution. Figure b related to figure c by a flop.}
\label{conifoldsingres}
\end{figure}

As a toric fan, we require each coordinate $z_i$ to correspond to a vector $w_i$ in $\Z^3$ which span $\C^3$. If we give a vector $w_i$ charge $Q_i$, corresponding to the relation $z_i\rightarrow \eta^{Q_i} z_i$, then the weighted sum of vectors is required to vanish as
\begin{equation}
Q^iw_i=0\:.
\end{equation}
The corresponding toric fan is thus given by the single {\it cone} spanned by the vectors
\begin{equation}
w_1=(1,1,0),\;\;w_2=(1,0,1),\;\;w_3=(1,0,0),\;\;w_4=(1,1,1)\:.
\end{equation}
Note that these vectors all lie in the same plane, which is true for all toric Calabi-Yau spaces. Hence three-dimensional toric Calabi-Yau geometries can be described by diagrams in a two-dimensional lattice. The diagram for the singular conifold is given in Figure \ref{conifoldsingres}. 
We see that the vectors spanning its fan are not linearly indepeendent and hence the corresponding variety is singular. To de-singularise it, giving the geometry $\tilde {\cal C}$, we can sub-divide the cone into two cones as shown in Figure \ref{conifoldsingres}. This can be done in two ways, corresponding to the two small resolutions of the conifold. The change going from one resolution to the other is the flop.

\subsubsection*{Toric quotients of the conifold}
\begin{figure}[H]
\begin{center}
\subfloat{\includegraphics[width = 1.6in]{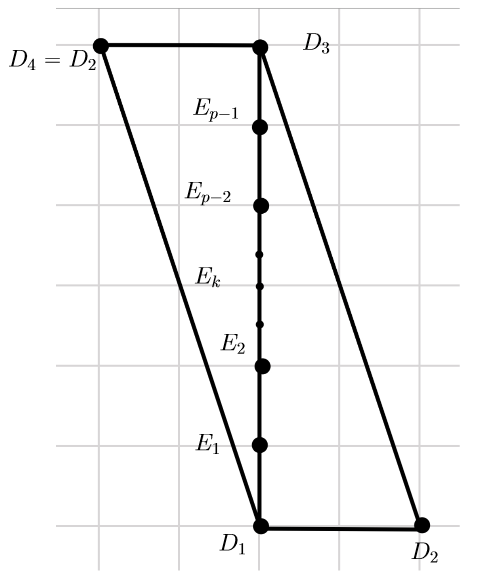}}
\caption{Diagram of the $\Z_p$-quotient of the conifold and its regular resolution.}
\label{fig:ToricConZp}
\end{center}
\end{figure}

\begin{figure}[H]
\begin{center}
\subfloat{\includegraphics[width = 3in]{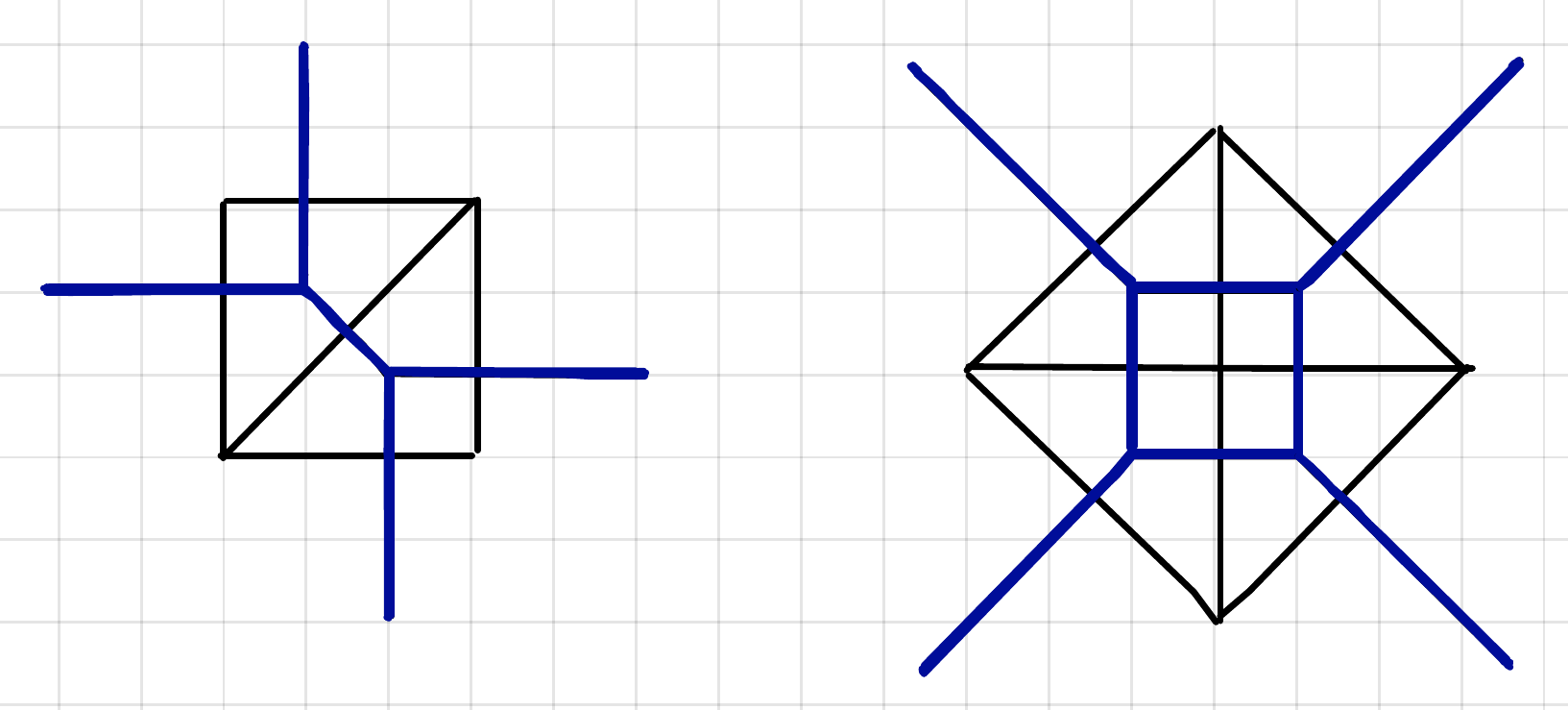}}
\caption{The conifold and its $\Z_2$-quotient and their dual diagrams in blue.}
\label{z2conifold}
\end{center}
\end{figure}

Next, we consider the toric description of the orbifolds of the conifold. Without going into too much detail, one can represent $\Z_p$-quotients of the conifold $\mathcal{C}_{p,k}$ as parallelograms in the lattice that only touch lattice points at it's four corners. The  $\mathbb{Z}_{p}$ action is defined through \cite{Davies:2013pna}
\begin{equation}
(y_{1},y_{2},y_{3},y_{4}) \sim (\lambda y_{1},\lambda^{k} y_{2},\lambda^{-k} y_{3},\lambda^{-1} y_{4}),
\end{equation}
where $\lambda = e^{2 \pi i / p}$, and the gcd$(p,k) = 1$. In the following we will only be concerned with the case where $k = 1$. In this case, the number of internal points of the parallelogram is $p-1$. The case of the $\Z_2$-quotient is drawn in Figure \ref{z2conifold}, along with its unique complete crepant resolution. The toric diagrams of the above described ${\widetilde{\cal C}}/\Gamma_{A_{p-1}}$ geometries are of this form, where all the internal points are stacked on top of each other. The vertices of the hyperconifold may be taken to be: $(1,0,0), (1,1,0),$ $ (1,0,p), (1,-1,p)$ in the toric lattice. This is drawn in Figure \ref{fig:ToricConZp}

The singularities of these hyperconifolds can be partially or completely removed by triangulating the diagram, which corresponds to a crepant resolution. The number of distinct triangulations depends on the parity of $p$. For even $p$ there is a unique triangulation. In contrast, for odd $p$ there are two triangulations. One of these triangulations matches the even case, where we call it the regular triangulation, while the other is obtained by a flop where the flop is performed using the $\P^{1}$ of the resolved conifold. The number of triangles needed to fully desingularise the space is equal to $ 2 \times p$. The number of distinct 2-cycles, which form a basis for the second homology, is equal to ${\rm dim }(H_{2}(\widetilde{\widetilde{\cal C}/\Gamma_{A_{p-1}}}) = p$.\\

Once a resolution of a singular toric geometry is established we can draw the {\it dual toric diagram}. The procedure for how this is done is straight forward, and we exemplify the procedure for the resolved conifold and its $\Z_2$-quotient in Figure \ref{z2conifold}. We see from this picture that aside from the four lines stretching to infinity (corresponding to non-inequivalent non-compact curves), there is also an exceptional compact set. For the resolved conifold, this is just a line of finite length, which in the geometry corresponds to a compact genus 0 curve. For the conifold mod $\Z_2$, the exceptional set is given by a square. In this case this is a compact divisor corresponding to $\mathbb{P}^1\times \mathbb{P}^1$, but could also be a Hirzebruch surface for more general cases we will consider.

Given a toric diagram and its resolution (triangularisation), we can work out the intersection numbers of the various compact and non-compact divisors involved in the geometry. First we assign a compact divisor $E_i$ to each internal non-boundary vertex of the triangulated diagram. These correspond to compact divisors of the exceptional set in the final geometry. Secondly, we assign a non-compact divisor $D_i$ to each vertex on the boundary of the diagram. If we denote by $e_1=(1,0,0)$, $e_2=(0,1,0)$ and $e_3=(0,0,1)$ the three unit vectors of the lattice, then we should also impose the three conditions
\begin{equation}
\label{eq:CondDivs}
\sum_{\rho}\langle e_i,v_{\rho}\rangle{\cal D}_{\rho}=0\:.
\end{equation}
where $v_\rho$ is the lattice vector corresponding to the divisor ${\cal D}_\rho$, and the sum is over all compact and non-compact divisors ${\cal D}_\rho$. The general rule for computing intersection numbers is that three divisors intersect once if they lie within a shared cone. Using \eqref{eq:CondDivs}, one can then compute triple intersections such as ${\cal D}_\rho^2{\cal D}_\kappa$ and ${\cal D}_\rho^3$, keeping in mind that at least one of the divisors should be compact for non-compact varieties. In the case of a hyperconifold  with the regular triangularisation $\tilde{\mathcal{C}}_{p}$ the non-zero triple intersection number are \cite{Closset:2018bjz}:
\begin{equation}\label{intersectionnumb}
        D_{2}E_{a}E_{b} = - A_{ab}, \; \ E_{a}^{3} = 8, \; \ E_{a-1}^{2}E_{a} = p - 2a, \; \ E_{a-1}E^{2}_{a} = 2 (a-1) - p , 
\end{equation}
where $A_{ab} = 2 \delta_{a,b} - \delta_{a, b+1} - \delta_{a+1,b}$ is the Cartan matrix of type $A_{p-1}$. Here $D_2$ is the leftmost/rightmost divisor on the boundary of the toric diagram. 

\begin{figure}[H]
\begin{center}
\subfloat{\includegraphics[width = 3in]{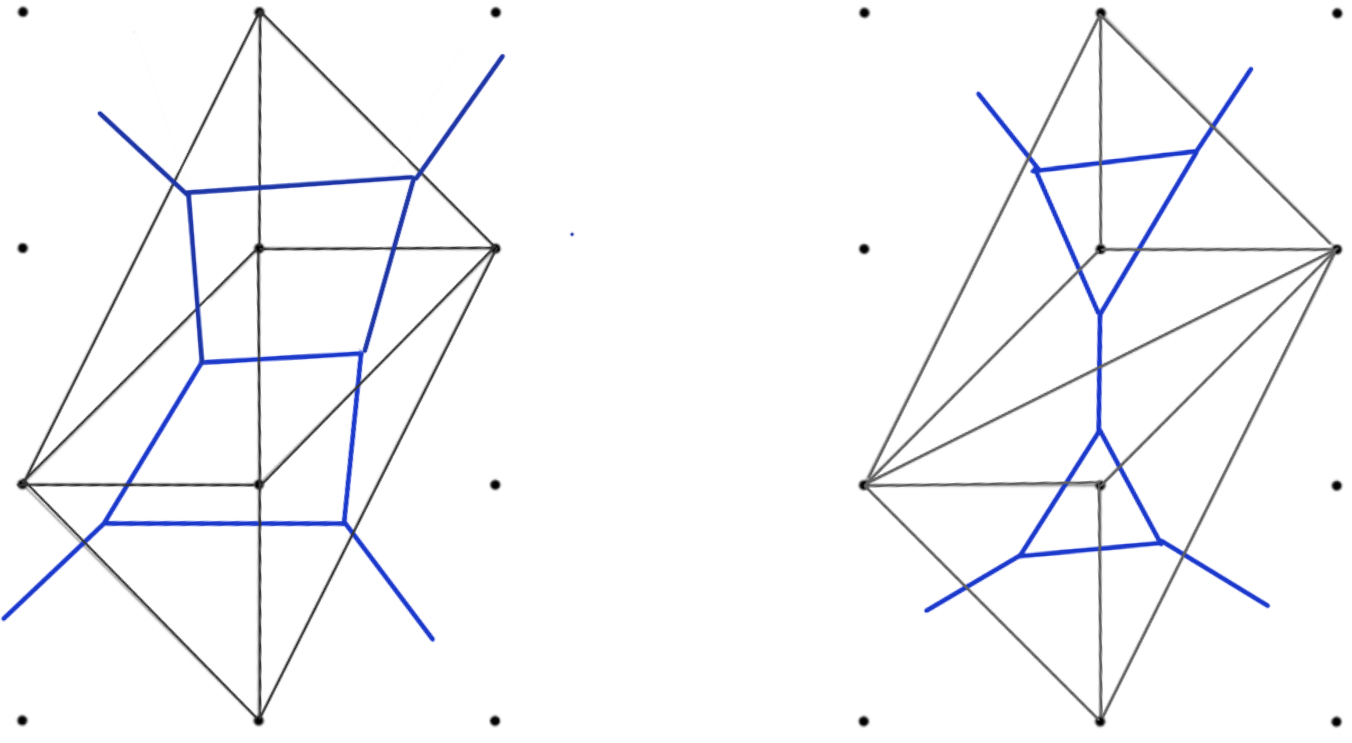}} \;\;\;\;\;
\caption{The toric $\Z_{3}$ example.}
\label{z3conifold}
\end{center}
\end{figure}

\subsubsection*{Example: $G_2$-holonomy spaces from the $\Z_2$-quotient}
To get a feel for how to use these rules, let us consider the hyperconifold $\widetilde{\cal C}_2$. Following Figure \ref{z2conifold} and using \eqref{eq:CondDivs}, we see that $D_3=D_1$, $D_4=D_2$ and $E=-2D_1-2D_2$. The non-vanishing triple intersections are
\begin{equation}
ED_1D_2=1,\;\; E^2D_1=E^2D_2=-2,\;\;E^3=8\:.
\end{equation}
Using this, we can apply the method of FHN for using such geometries to construct local $G_2$ manifolds by considering circle bundles over Calabi-Yau manifolds \cite{foscolo2017complete}. As noted above, the crucial condition for this to be possible is that
\begin{equation}
\label{eq:G2cond}
c_1\wedge\omega=0\in H^4(X)\:,
\end{equation}
where $X$ is the non-compact Calabi-Yau threefold, $\omega$ is its K\"ahler class and $c_1$ is the first Chern class of the circle bundle involved. 
This condition is equivalent to $c_1\wedge\omega$ integrating to zero over compact divisors in $X$, i.e.
\begin{equation}\label{eq:G2cond:II}
\int_{E_i}c_1\wedge\omega=c_1\cdot\omega\cdot E_i = 0\;\;\;\forall\;E_i\:,
\end{equation}
where $E_i$ corresponds to a compact divisor in $H_4(X)$. 

Let us expand the K\"ahler class and $c_1$ in a complete set of divisors as
\begin{align}
\omega&=t_1D_1+t_2D_2\\
c_1&=\kappa_1D_1+\kappa_2D_2\:.
\end{align}
By inspection of the dual diagram the geometry has two compact curves, $C_1=D_1E$ and $C_2=D_2E$.
Both curves are genus zero $\mathbb{P}^1$ curves. Quantisation conditions then require that
\begin{equation}
\int_{ C_i}c_1=c_1\cdot C_i=\kappa_i
\end{equation}
are integers. The volume of the curve $ C_i$ is furthermore given by $t_i$. There is a unique compact divisor $E$, so \eqref{eq:G2cond:II} gives
\begin{equation}
\label{eq:Z2cond}
c_1\cdot \omega\cdot E=(\kappa_1t_2+\kappa_2t_1)\, D_1\cdot D_2\cdot E=\kappa_1t_2+\kappa_2t_1=0\:.
\end{equation}
Modulo conventions, this is exactly the condition used in \cite{foscolo2018infinitely} to construct infinitely many new families of complete cohomogeneity one $G_2$-manifolds. 

\subsubsection*{Example: new $G_2$-manifolds from the $\Z_3$-Quotient}
\label{sec:ConZ3}
The next example we consider is the $\Z_3$-quotient of the conifold. It's toric diagram together with its two smooth crepant resolutions is also given in Figure \ref{z3conifold}, with the corresponding dual diagrams. In this case, the geometry has four non-compact divisors $D_i$ and two compact divisors $E_i$, satisfying the following relations
\begin{align}
E_1+E_2+D_1+D_2+D_3+D_4&=0\\
D_2&=D_4\\
D_3&=D_2+E_1+2D_1\:.
\end{align}
Both resolutions have two compact divisors. In the regular resolution the exceptional set is given by a union of two Hirzebruch surfaces $F_1$, both fibered over a common $\mathbb{P}^1$ base. The second resolution has as its compact set two $\mathbb{P}^2$ surfaces, both intersecting a $\mathbb{P}^1$ at separate points.

{\bf Type I resolution:} It is easy to work out the intersection numbers for the divisors of the two different resolutions. Using all this, we can work out the $G_2$ condition \eqref{eq:G2cond} for both cases. Let us begin with the first case. We expand the K\"ahler class and first Chern class as\footnote{In these examples it is convenient to use the non-compact divisors as a basis. For higher quotients we will also need to include exceptional divisors in the expansion.}
\begin{align}
\omega&=t_1D_1+t_2D_2+t_3D_3\\
c_1 &=\kappa_1D_1+\kappa_2D_2+\kappa_3D_3\:.
\end{align}
The exceptional set has three compact genus zero curves, $ C_1=D_2E_1$, $ C_2=E_1E_2$ and $ C_3=D_2E_2$. These curves have volume $t_1>0$, $t_2>0$ and $t_3>0$ respectively. Integrating $c_1$ over $ C_i$, we further find that $\kappa_i$ should be an integer. We then find
\begin{equation*}
\omega\cdot c_1=(t_1\kappa_1+\kappa_1t_2+\kappa_2t_1)D_1D_2+(t_3\kappa_3+\kappa_3t_2+\kappa_2t_3)D_3D_2\:,
\end{equation*}
where we have used relations such as $D_1^2=D_1D_2$, $D_3^2=D_3D_2$ and $D_2^2=0$ which can be worked out from the intersection relations. We thus get the two conditions
\begin{align}
\label{eq:CondZ31}
t_1\kappa_1+\kappa_1t_2+\kappa_2t_1&=0\\
\label{eq:CondZ32}
t_3\kappa_3+\kappa_3t_2+\kappa_2t_3&=0\:.
\end{align}
We can solve these equations for $t_1$ and $t_3$ in terms of $t_2$, giving
\begin{equation}
t_1=-\frac{\kappa_1t_2}{\kappa_1+\kappa_2}\:,\;\;\;t_3=-\frac{\kappa_3t_2}{\kappa_3+\kappa_2}\:.
\end{equation}
In order for these solutions to make sense, with both $t_1$ and $t_3$ non-negative, we must require that $\kappa_2$ has opposite sign of $\kappa_1$ and $\kappa_3$, and $\vert \kappa_2\vert\ge \vert \kappa_{1,3}\vert$. In the case of $\vert \kappa_2\vert = \vert \kappa_{1,3}\vert$, we see that the corresponding $t_{1,3}$ are left undetermined, while $t_2=0$ unless $\vert \kappa_2\vert = \vert \kappa_{1,3}\vert=0$. Note also that in this case for $\kappa_2\neq0$ we would have flux on a vanishing cycle $t_2=0$, the meaning of which is not completely clear. We will come back to this in section \ref{sec:OpneClosed}. 

{\bf Type II resolution:} For completeness we also consider the second type of resolution, whose dual diagram is depicted in Figure \ref{z3conifold}. This case is perhaps less relevant from a physical point of view, as the interpretation in terms of four-dimensional ${\cal N}=1$ gauge theory is less obvious. 

We still have $D_1^2=D_1D_2$ and $D_3^2=D_3D_2$, but now $D_2^2\neq0$. We denote 
$\tilde C_2=D_2^2$ and $\tilde C_1=D_2E_1$ and $\tilde C_3=D_2E_2$. Let us also expand the corresponding K\"ahler and first Chern class as
\begin{align}
\omega&=u_1D_1+u_2D_2+u_3D_3\\
c_1 &=a_1D_1+a_2D_2+a_3D_3\:.
\end{align}
The condition \eqref{eq:G2cond} now gives
\begin{equation}
c_1\cdot\omega=(u_1a_1+u_1a_2+u_2a_1)D_1D_2+(u_3a_3+u_3a_2+u_2a_3)D_3D_2+u_2a_2D_2^2=0\:.
\end{equation}
In particular, computing $E_i\cdot c_1\cdot\omega$ we find
\begin{align}
(u_1+u_2)(a_1+a_2)&=0\\
(u_3+u_2)(a_3+a_2)&=0\:.
\end{align}
However, it is easy to check that the volume of $\tilde C_1$ and $\tilde C_3$ are given by $u_1+u_2$ and $u_3+u_2$ respectively. A similar computation also fixes $a_1+a_2=\kappa_1$ and $a_3+a_2=\kappa_3$ to be integers. We must therefore require $\kappa_1=\kappa_3=0$ if we want a smooth solution. For this choice the volumes of the curves are unconstrained, while in the regular resolution, given an allowable flux configuration, $\omega$ is fixed up to scale.

\subsection{Interpretations in Physics}
Compactifying type IIA string theory on non-compact toric geometries gives rise to ${\cal N}=2$ supersymmetric theories in four dimensions, in the limit of decoupled gravity. Depending on the corresponding toric resolution, some of these theories have an interpretation as non-Abelian gauge theories in various phases. Consider for example the regular resolution of the $\Z_3$-quotient of the resolved conifold above. This particular example corresponds to a resolution of the $\widetilde{\cal C}/\Gamma_{A_{2}}$ geometry, and has a gauge theory interpretation. Sending the size of the curves $ C_1$ and $ C_3$ to zero then gives rise to an $A_2$ singularity in the $\R^4/\Z_3$ fiber. This can be interpreted as an $SU(3)$ gauge theory in an unbroken phase, whose coupling constant is given by the inverse volume of the curve $ C_2$. Giving a volume to the curves $ C_1$ and $ C_3$ is then equivalent to Higgsing the gauge theory, breaking it to a Coulomb phase. It is less clear how the Type II resolution fits into such a gauge theory interpretation.

The ${\cal N}=2$ theory may further be explicitly broken to an ${\cal N}=1$ theory by the introduction of a superpotential, induced by the presence of RR two-form flux which classically reads
\begin{equation}
\label{eq:ClassW}
W^c=\tfrac12\int_Xc_1\wedge T\wedge T\:,
\end{equation}
 where $c_1$ denotes the fist Chern-class of the line-bundle associated with the fluxes and 
 \begin{equation}
T=\alpha^iE_i+\mu^\alpha D_\alpha=T^\mu\hat E_\mu\:,
\end{equation}
where $T^\mu=\{\alpha^i,\mu^\alpha\}$, and $\{\hat E_\mu\}$ now denotes a sub-set of all divisors $\{{\cal D}_\rho\}$, subject to the identifications \eqref{eq:CondDivs}. Note that in the conifold case we can take the set of non-compact divisors to consist of a single divisor $D=D_2$.\footnote{For more general non-compact Calabi-Yau manifolds, there could in principle be more non-compact divisors to consider.} Indeed, we have three relations between the divisors, enabling us to remove three non-compact divisors of our choosing. The complexified K\"ahler moduli are $\alpha^j=v^j+i\,a^j$, where $a^j$ are the axions, and
\begin{equation}
\omega=v^iE_i+{\rm Re}(\mu)D\:.
\end{equation}
In the limit of decoupled gravity, a supersymmetric vacuum requires that $\delta_T W^c=0$, where $\delta T$ corresponds to a physical field. That is $\delta T$ has compact support. It follows that we should require
\begin{equation}
\delta T=\delta\alpha^iE_i\:.
\end{equation}
The condition we hence obtain for a supersymmetric vacuum reads
\begin{equation}
\label{eq:FHNcondition}
\int_{E_i}c_1\wedge T=0\;\;\;\Leftrightarrow\;\;\;c_1\cdot T\cdot E_i=0\;\;\forall\;E_i\:.
\end{equation}
In particular, we must have $c_1\cdot\omega\cdot E_i=0$ for all $E_i$, which is precisely the requirement of \cite{foscolo2017complete} to be able to construct a complete $G_2$ metric on this line bundle.  

The classical superpotential \eqref{eq:ClassW} receives a perturbative quantum correction at one-loop, in addition to non-perturbative quantum corrections, often called instanton corrections. Indeed, we may write the classical superpotential as \cite{Vafa:2000wi}
\begin{equation}
W^c=N^\mu\partial_\mu{\cal F}_0^c(T)\:,
\end{equation}
where ${\cal F}_0^c(T)$ denotes the classical pre-potential\footnote{This is the classical prepotential of thee ${\cal N}=2$ theory one gets from compactifying type IIA on the Clabi-Yau $X$. It is the full one-loop exact prepotential of the five dimensional ${\cal N}=1$ theory one gets from compactifying $M$-theory on $X$ \cite{Intriligator:1997pq}.}
\begin{equation}
{\cal F}_0^c(T)=-\tfrac16\int_X T\wedge T\wedge T
\end{equation}
where we have expanded 
\begin{equation}
c_1=N^\mu\hat E_\mu=n^i E_i+m^\alpha D_\alpha\:.
\end{equation}
The flux parameters $\{n^i,m^\alpha\}$ will satisfy certain integral constraint coming from quantisation conditions imposed on $c_1$ when integrated over compact curves in the geometry. It is the pre-potential that reccives quantum corrections, giving the full quantum potential ${\cal F}_0$. The quantum corrected superpotential then becomes
\begin{equation}
W = N^\mu\partial_\mu{\cal F}_0(T)\:.
\end{equation}
For toric geometries, methods have been developed to compute ${\cal F}_0$ exactly through the topological vertex \cite{Iqbal:2002we, Aganagic:2003db}.

\subsubsection*{Example: $\Z_2$-quotient of the resolved conifold}
Let us apply this to the resolvedd conifold mod $\Z_2$ we considered above. We can compute the classical pre-potential and superpotential. If we write 
\begin{equation}
T=\alpha\, E+\mu\, D_2\:,
\end{equation}
we get
\begin{equation}
{\cal F}_0^c(T)=-\tfrac43\alpha^3+\mu\alpha^2+\alpha{\rm-independent}\:.
\end{equation}
This corresponds to the classical pre-potential of the Coulomb branch of a $SU(2)$ gauge theory, where we identify $\mu$ with the inverse complexified bare coupling constant and $\phi=-\alpha$ with the the Coulomb branch scalar \cite{Intriligator:1997pq, Xie:2017pfl, Closset:2018bjz}.

Writing $c_1=\kappa_0D_2+\kappa_1D_1$ we also get the classical superpotential as
\begin{equation}
W^c=\kappa_1\mu\alpha-(\kappa_0+\kappa_1)\alpha^2+\alpha{\rm-independent}\:.
\end{equation}
In the notation of the $\Z_2$-example above, we have
\begin{align}
t_2&=\mu-2\alpha\\
t_1&=-2\alpha\:,
\end{align}
where we see that the volume of a curve, ``the fiber", corresponds to one of the fields while the inverse coupling $\mu$ is a linear combination of the volumes of both curves. Note that classical supersymmetric vacua should satisfy
\begin{equation}
\label{eq:Z2SuperpotCond}
\partial_\alpha W^c=0=\kappa_1t_2+\kappa_2t_1\:,
\end{equation}
as it should be. Note that the condition that this vanish for supersymmetric solutions forces the two integers $\kappa_i$ to be of opposite sign.

\subsubsection*{Example: $\Z_3$-quotient of the resolved conifold}
Let us also consider the regular resolution of the $\Z_3$-quotient of the resolved conifold. Let us work with the parametrisation of the flux and the complexified K\"ahler two-form as : 
\begin{equation}
\begin{split}
    c_1 &= \kappa_{1} D_{1} + \kappa_{2} D_{2} + \kappa_{3} D_{3},
\\
    T &= \alpha_{1} E_{1} + \alpha_{2} E_{2} + \mu D_{2}
\end{split}
\end{equation}
We compute the prepotential
\begin{equation}
    \mathcal{F}^{c}(\mu, \alpha_{i}) = \frac{4}{3} \alpha_{1}^{3} - \frac{4}{3}  \alpha_{2}^{3} + \frac{4}{3}  \alpha_{1}\alpha_{2}\mu - \alpha_{1}^{2}\mu - \alpha_{2}^{2} \mu - \frac{1}{2} \alpha_{1}\alpha_{2}^{2} - \frac{1}{2} \alpha_{1}^{2}\alpha_{2}\:,
\end{equation}
and the superpotential is given by
\begin{equation}
    W^{c}(\kappa_{i},\alpha_{i}, \mu) = -\frac{\alpha_{1}^{2}}{2} (3 \kappa_{1} + 2 \kappa_{2}) - \frac{\alpha_{2}^{2}}{2} (2 \kappa_{2} + 3 \kappa_{3}) + \alpha_{1} \mu \kappa_{1} + \alpha_{2} \mu \kappa_{3} + \alpha_{1} \alpha_{2} \kappa_{2}\:..
\end{equation}
The coupling $\mu$ is then given in terms of the volumes $t_i$ of thecurves as one set $T = \alpha_{1} E_{1} + \alpha_{2} E_{2} + \mu D_{2} = t_{1} D_{1} + t_{2} D_{2} + t_{3} D_{3}$, then 
\begin{equation}
 \mu \sim t_{1} - t_{2} + t_{3}\:.
\end{equation}
The fields are given as 
\begin{equation}
\alpha_{1} = -\tfrac{1}{3}(2t_{1} + t_{3}),\;\;\;\alpha_{2} = -\tfrac{1}{3}(2t_{3} + t_{1})\:.
\end{equation}
The $G_{2}$-condition is determined by $\partial_{\alpha_{i}}W = 0;\  i = 1,2$. Thus we have
\begin{equation}
\begin{split}
\partial_{\alpha_{1}}W &= 0 = \kappa_{1} (\mu - 3 \alpha_{1}) + \kappa_{2}(\alpha_{2} - 2\alpha_{1})
\\
\partial_{\alpha_{2}}W &= 0 = \kappa_{3} (\mu - 3 \alpha_{2}) + \kappa_{2}(\alpha_{1} - 2\alpha_{2})\:.
\end{split}
\end{equation}
With some algebra one arrives at equations \eqref{eq:CondZ31}-\eqref{eq:CondZ32}.

\subsection{The general case}
\begin{figure}[H]
\begin{center}
\subfloat{\includegraphics[width = 5in]{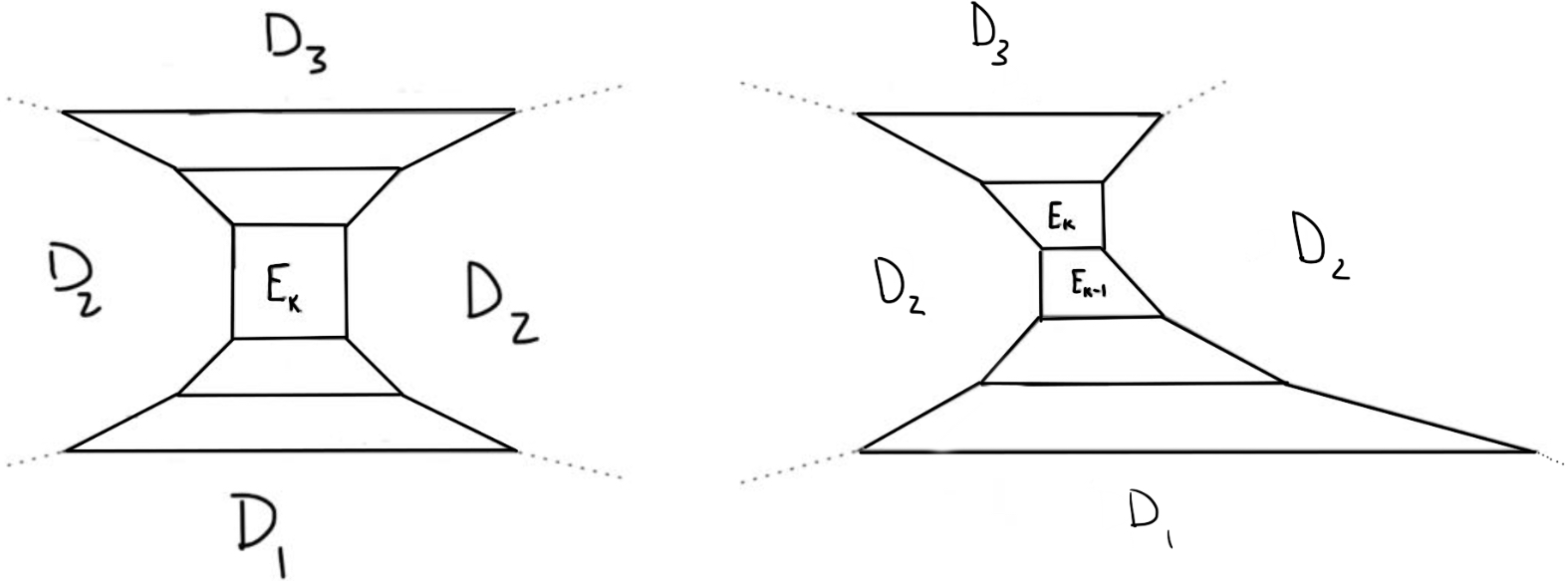}}
\caption{The figures depicted the general dual diagrams of the regular resolution of $\Z_p$-the quotient of the Hyperconifold. The figure on the left is the case of even p. On the right is the odd-p case.}
\label{fig:DualConZp}
\end{center}
\end{figure}

For general $\Z_p$ quotients, of which we draw the toric dual diagram in Figure \ref{fig:DualConZp}, we need a more systematic way to distinguish the divisor corresponding to the gauge coupling, and the fields of the theory. From a physical perspective, it is natural to associate the divisors corresponding to the volume forms of the $\mathbb{P}^1$'s living in the resolved fiber $\widetilde{\R^4/\Z_p}$ as the fields of the theory. These are the compact exceptional divisors $E_i$. In the UV theory these are all shrunk to zero size leaving a two-sphere, ``the base". This is also the ``two-sphere at infinity". The volume form of this curve is dual to a divisor which we should naturally associate with the bare gauge coupling. We identify this divisor as follows.

Begin by constructing the $(p+3)\times 3$ matrix $M_p$ whose columns consist of the $\Z^3$ coordinates of the divisors in the toric diagram. $(p+3)$-vectors in the kernel of this matrix then correspond to curves in the geometry in the following sense: given $V\in\ker(M_p)$ we can define the divisor
\begin{equation}
D_V=\sum_{\mu=1}^{p+3} a_\mu\hat E_\mu\:,
\end{equation}
where $a_\mu$ are the components of $V$. Then $D_V$ is the volume form of a particular (sum of) curves in the geometry.\\ 

Going to the UV, or shrinking the fibers to zero size leaves only one two-sphere in the geometry. This is the sphere at infinity, $S^2_\infty$, which is naturally associated with the inverse of the UV coupling. We therefore expect a unique vector $V_\infty$ in the kernel of $M_p$ which only consists of non-compact divisors, and which corresponds to the volume form of this curve. This is indeed the case, and the corresponding divisor is
\begin{equation}
D_\infty=D_1-D_2+D_3-D_4=D_1+D_3-2D_2\:,
\end{equation}
as $D_2=D_4$ in all the geometries under consideration. 

There is also a method for evaluating the flux of the $M$-theory circle bundle over the two-sphere at infinity $S^2_\infty$. For $1\le i\le p-1$ we define the curves
\begin{equation}
 C_i=D_2\cdot E_i\:,
\end{equation}
where $D_2$ is the divisor corresponding to the point furthest to the left/right in Figure \ref{fig:ToricConZp}. In addition we define a distinguished curve
\begin{equation}
 C_0=E_k\cdot E_{k+1}\:,
\end{equation}
where $k$ is taken as close to $p/2$ as possible. This curve, ``the base",  is in close relation with the two-sphere at infinity. The curves $ C_\mu$ generate the Mori cone of the toric diagram. We then write
\begin{equation}
S^2_\infty = \sum_{i=0}^{p-1} p_i C_i\:.
\end{equation}
The numbers $p_i$ are then determined up to an overall constant by requiring that
\begin{equation}
S^2_\infty\cdot D_{V_j}=0
\end{equation}
for all $V_j$ orthogonal to $V_\infty$. The overall constant can be determined by requiring $p_0=1$, as $S^2_\infty$ and $ C_0$ should agree at infinity. In terms of the curves $ C_i$ one finds for $p$ even
\begin{equation}
\label{eq:S2infpeven}
S^2_\infty= C_0- C_1-2\, C_2-..-\tfrac{p}{2}\, C_{p/2}-..-2\, C_{p-2}- C_{p-1}\:,
\end{equation}
and 
\begin{equation}
\label{eq:S2infpodd}
S^2_\infty= C_0- C_1-2\, C_2-..-\tfrac{p-1}{2}\, C_{(p-1)/2}-\tfrac{p-1}{2}\, C_{(p+1)/2}-..-2\, C_{p-2}- C_{p-1}\:,
\end{equation}
for $p$ odd. The charge $Q$ of the $M$-theory circle bundle over this two-sphere is then given by
\begin{equation}
Q=c_1\cdot S^2_\infty\:,
\end{equation}
which gives a weighted sum of the charge integers $\kappa_i$ of the circle bundle associated to each curve $ C_i$, in accordance with \eqref{eq:S2infpeven} for $p$ even and \eqref{eq:S2infpodd} for $p$ odd. \\

If we choose a parametrisation of the complexified K\"ahler form as
\begin{equation}
T=\mu\,D_2 +\alpha^iE_i\:,
\end{equation}
it can be shown that the geometric prepotential in this parametrisation agrees with the one-loop Coulomb-branch prepotential of the dual five-dimensional ${\cal N}=1$ gauge theory (derived from $M$-theory compactified on the local Calabi-Yau) \cite{Closset:2018bjz}, 
\begin{equation}
\mathcal{F}^{c}_{0}(\mu, \alpha_{i}) = \frac{4}{3} \sum_{i=1}^{p-1} \alpha_{i}^{3} + \sum_{i=1}^{p-1} (p-2i) [\alpha_{i-1}^{2}\alpha_{i} - \alpha_{i+1}^{2}\alpha_{i}] - \frac{1}{2} \mu \sum_{i,j = 1}^{p-1} A^{ij}\alpha_{i}\alpha_{j} \ ,
\end{equation}
which also agrees with the classical prepotential of the four-dimensional ${\cal N}=2$ theory if we identify $\mu$ with the inverse of the complexified bare gauge coupling and $\alpha^i$ with the fields of the theory. The effective coupling matrix element, defined as
\begin{equation}
\tau_{ij} := \frac{\partial^{2}}{\partial\alpha^{i}\partial\alpha^{j}} \mathcal{F}^{c}_{0}(\mu, \alpha_{i})
\end{equation} 
is then equal to
\begin{equation}\label{eq:effectivecoupling}
\tau_{ij} = 8 \ \alpha_{i}\ \delta_{ij} - \mu A_{ij}\:.
\end{equation}
A short computation also gives
\begin{equation}
S^2_\infty\cdot T=\mu\:,
\end{equation}
and so the volume of the sphere at infinity corresponds to the bare gauge coupling as expected.\\ 

Supersymmetric vacua then correspond to solutions of \eqref{eq:FHNcondition}. Additionally, for physically consistent solutions we must also require that the volumes of curves and divisors in the geometry are non-negative. If we write $c_1$ and $T$ in terms of $c_1\cdot C_i=\kappa_i\in\Z$ and $T\cdot C_i=t_i\ge0$ we find that the constraint
\begin{equation}
c_1\wedge\omega=0
\end{equation}
is equivalent to the system of $p-1$ linear equations
\begin{equation}
\label{eq:SusyConstrLin}
t_0 \ \kappa_i+t_i \ c_1\cdot C_0(i)=0\:,
\end{equation}
where $C_0(i)$ are the curves corresponding to the horizontal edges in the dual diagrams of Figure \ref{fig:DualConZp}. One can show that $c_1\cdot C_0(i)$ is a linear combination of the flux integers $\kappa_j$ for $j$ between $i$ and $\tfrac{p}{2}\pm 1$. Equations \eqref{eq:SusyConstrLin} can only be satisfied with $t_0,..,t_{p-1}\ge0$ if (up to an overall change of sign) $\kappa_i\ge0$ for $i=1,..,p-1$ and $c_1\cdot C_0-(i)\le 0$  for $i=1,..,p-1$. As in the $\Z_3$ case discussed explicitly earlier, a computation shows that the latter inequalities force $\kappa_0<0$ and constrain the size of $\kappa_i$,  $i=1,..,p-1$, in terms of the size of $\vert\kappa_0\vert$ and the $\kappa_j$ with $j$ between $i$ and $\tfrac{p}{2}\pm 1$.

\section{Open-Closed Duality}
\label{sec:OpneClosed}
The closed type IIA string geometries described above correspond to the infra-red theories of configurations of D-branes wrapped on quotients of the deformed conifold. To see why, consider the case of the $\Z_2$-quotient of the resolved conifold, with $N=\kappa_0-\kappa_1$ units of flux on the sphere at infinity. For the case $\kappa_{1} = 0$, we must have $t_F=0$ with $t_B$ arbitrary by equation \eqref{eq:Z2SuperpotCond}. The geometry describes an ${\Rr}^4/\Z_2$ fibered over a $\Pp^1$ with $N$ units of flux. We can lift this to $M$-theory as a $\Z_2\times\Z_N$-quotient of the Bryant-Salamon manifold $\Rr^4\times S^3$ \cite{bryant1989construction}. If we now do a $G2$ flop, reversing the action of $\Z_2\times\Z_N$ so that $\Z_2$ acts on $ S^3$ while $\Z_{N}$ acts on $\Rr^4$, we can make a type IIA reduction on the Hopf-fiber of $\Rr^4$. The resulting type IIA geometry is then given by $N$ D-branes wrapped on a $\Z_2$-quotient of the deformed conifold, which effectively gives rise to an $SU(N)$ gauge theory on $ S^3/\Z_2$. This situation generalises to more complicated flux configurations and quotients of the conifold as we will now describe. \\

Let's return to consider the geometries $X_{\kappa_0,\kappa_1,..,\kappa_{p-1};q}$. At infinity these geometries are circle fibrations over $ S^3/\Z_p\times  S^2$ with $N=-Q$ units of flux on $ S^2$, where as we have seen
\begin{equation}
N=-\kappa_0+\kappa_1+2\kappa_2+..+2\kappa_{p-2}+\kappa_{p-1}\:,
\end{equation}
while $q$ determines the monodromy $\Z_p\rightarrow U(1)$ of the bundle on $ S_1^3/\Z_p$. In order to calculate $q$ in terms of the fluxes, recall that the resolved space $\widetilde{\widetilde{\cal C}/\Z_p}$ is a fibration over $\P^1$ with fiber $\widetilde{\R^4/\Z_p}$. Consider the complex line bundle $L_i$ with first Chern class the Poincare dual of the curve $C_i$. The McKay correspondence implies that the restriction of $L_i$ to the $ S^3_1/\Z_p$ at infinity of each fibre is the flat line bundle $( S^3_1\times\C)/\Z_p$ where $\Z_p$ acts on $\C$ via the representation of weight $i$. Therefore we must have
\begin{equation}
q=\kappa_1+2\kappa_2+..+(p-1)\kappa_{p-1}\;\;\textrm{mod}\;p\:.
\end{equation}
The space at infinity is given as
\begin{equation}
M_{(p,N,q)}=( S_1^3\times S_2^3)\Big/\Gamma_{(p,N,q)}\:.
\end{equation}
From this perspective, the $M$-theory flop is a matter of ``filling in" either $ S_1^3$ (closed/IR side) or $ S_2^3$ (open/UV side). In particular, in the UV for $q=0$ we have $\Gamma_{(p,N,0)}=\Z_p\times\Z_N$. In this case, reducing $M$-theory on $\R^4/{\Z}_N$ results in $N$ $D6$ branes wrapping the Lens space $ S^3/\Z_p$ and giving rise to an effective $SU(N)$ theory on this space.\\ 

For $q\neq0$ things are a bit more subtle as the group $\Gamma_{(p,N,q)}$ does not factor anymore. However, it is still a group of order $N\times p$. E.g. if we set $q=1$ we see that the group in question $\Gamma_{(p,N,1)}\cong \Z_{pN}$. This group has a $\Z_N$ sub-group which acts trivially on $ S^3_2$ given by elements of the form $\{(1,1),(1,\kappa^p,),(1,\kappa^{2p}),..,$ $(1,\kappa^{(N-1)p})\}$ where $\kappa$ is the generator of $\Z_{pN}$. The other elements act freely. This sub-group hence produces an $SU(N)$ gauge theory on $ S^3_2$, and factoring out by the remaining elements of $\Gamma_{(p,N,1)}$ gives an $SU(N)$ theory on $ S^3/\Z_p$. It can similarly be checked that for generic $q$ there is a $\Z_N$ subgroup with fixed point the origin of $\R^4$ times $ S^3_2$, with the other elements acting freely. The result is hence an $SU(N)$ gauge theory on $ S^3/\Z_p$ for each $q$. This is similar to an observation of Witten for $M$-theory on similar quotients of $\R^4 \times  S^1$ \cite{Witten:1997kz}.\\

As we have seen, the monodromy $q$ is related to the flux integers $\{\kappa_1,..,\kappa_{p-1}\}$ associated to the resolved curves of the fiber $\widetilde{\R^4/\Z_p}$ and for each $q$ there is also a corresponding $SU(N)_q$ theory with a plethora of flat connections. Such a flat connection is determined through a flat $U(N)$ connection
\begin{equation}
\tilde g=(\lambda^0I_{{n}_0},\lambda I_{{n}_{1}},\lambda^2 I_{{n}_{2}},..,\lambda^{p-1}I_{{n}_{p-1}})
\end{equation}
with determinant $\lambda^q$ where
\begin{equation}
q={n}_1+2{n}_2+..+(p-1){n}_{p-1}\;\;{\rm mod}\; p\:.
\end{equation}
Here, $\lambda$ is the generator of $\Z_p$ and $\{{n}_0,{n}_1,..,{n}_{p-1}\}$ correspond to a decomposition of $N$ as
\begin{equation}
N={n}_0+{n}_1+..+{n}_{p-1}\:.
\end{equation}
The flat $SU(N)_q$ connection is then given as $g=\lambda^{\kappa_q}\tilde g$, where $\kappa_q$ is chosen so that $\det(g)=1$. This can be done if and only if $N$ is coprime to $p$. We then propose to identify the fluxes with the integers ${n}_i$ by
\begin{equation}
\kappa_i={n}_i\;\;\;i<p\:,\;\;\;{n}_0=N-(\kappa_1+..+\kappa_{p-1})\:.
\end{equation}

\subsection{The $\Z_2$-quotient of the conifold}
Let us consider the infinite number of $G_2$-holonomy manifolds $X_{(\kappa_0,\kappa_1)}$ of \cite{foscolo2018infinitely}, i.e. the $M$-theory lifts of the resolved $\Z_2$-quotients of the conifold now with the fluxes distributed more generally. We ask what the interpretation of these geometries on the deformed side, i.e. as configurations of D-branes wrapped on an appropriate $\Z_2$-quotient of the deformed conifold? \\

There are two cases, $q=0$  where $\Gamma_{(2,N,1)}=\Z_2\times\Z_{N}$ and $q=1$ where $\Gamma_{(2,N,1)}=\Z_{2N}$. Let us also assume that $N$ and $p$ are coprime. Then the different choices of flux integers $N=\kappa_0-\kappa_1$ match the different configurations of flat connections within two $SU(N)$ theories. 
The total number of such flat connections is \cite{Friedmann:2002ct}
\begin{equation}
{\rm Number\:of\:flat\:connections}=p\times\frac{(N+p-1)!}{N!p!}=N+1\:,
\end{equation}
when $p=2$. Note that this exactly matches the of ways to partition $N$ into $p$ integers, even for general $p$ and $N$.  

\begin{example}[$N=5$] Consider for example $N=5$. The matching of fluxes to flat connections then becomes 
\begin{center}
\begin{tabular}{ c|c|clc } 
 $(n_{0},n_{1})$ & $SU(5)_0$ & $SU(5)_1$ & Broken gauge group \\ 
 \hline
 $(5,0)$ & $(+++++)$ & - & $SU(5)$ \\ 
 $(4,1)$ & -  & $(+----)$ & $SU(4)\times U(1)$ \\ 
 $(3,2)$ & $(+++--)$ & - & $SU(3)\times SU(2)\times U(1)$ \\ 
 $(2,3)$ & - & $(+++--)$ & $SU(3)\times SU(2)\times U(1)$ \\ 
 $(1,4)$ & $(+----)$ & - & $SU(4)\times U(1)$\\ 
 $(0,5)$ & - & $(+++++)$ & $SU(5)$ \\ 
\end{tabular}
\end{center}
where $SU(5)_{0/1}$ corresponds to the two different gauge theories. 
\end{example}

\subsection{The $\Z_3$-quotient of the conifold}
Let us now see what happens when we consider $\Z_3$-quotients of the conifold, i.e. $p=3$ in the above. In this case the total number of flat $SU(N)$ connections on $ S^3/\Z_3$ is
\begin{equation}
{\rm Number\:of\:flat\:connections}=p\times\frac{(N+p-1)!}{N!p!}=\tfrac12(N+2)(N+1)\:.
\end{equation}
These are the triangle numbers for $N+1$, which exactly matches the number of ways to partition $N$ into three distinct integers. For example, if $N=4$ there are $3\times5=15$ flat connections, corresponding to the fifth triangle number.

\begin{example}[$N=4$] To get a feel for how the correspondence between flat connections and fluxes works, let us consider the example of $N=4$. From the open string perspective, this corresponds to three different $SU(4)_q$ theories on $ S^3/\Z_3$ parameterised by $q\in\{0,1,2\}$. We can decompose $N$ into three positive integers
\begin{equation}
N={n}_0 + {n}_1 + {n}_2\:.
\end{equation}
Note that this number precisely matches the number of flat connections for the three $SU(N)_q$ theories, which is fifteen for $N=4$. The procedure to match these integers to the flux integers $\kappa_i$ is then the following. We then match 
\begin{equation}
\kappa_0=-{n}_0\:,\;\;\kappa_1={n}_1\:,\;\;\kappa_2={n}_2\:.
\end{equation}
The corresponding flat $SU(N)$ connections are then given by considering
\begin{equation}
\tilde g=(\lambda^0I_{{n}_0},\lambda I_{{n}_{1}},\bar\lambda I_{{n}_{2}})\:,
\end{equation}
with $\lambda$ is a third root of unity and $\bar\lambda=\lambda^2$. Note that $\tilde g$ has determinant ${\rm det}(\tilde g)=\lambda^q$ where
\begin{equation}
q={n}_{1}+2\,{n}_{2}\;\; {\rm mod}\;\;3\:.
\end{equation}
The flat $SU(N)$ connection then takes the form
\begin{equation}
g=\lambda^{\kappa_q}\tilde g\:,
\end{equation}
where $\kappa_q$ is chosen so that $g$ has unit determinant. This can be done provided $N$ is coprime to $p=3$.\\

For $N=4$ we have the following table
\begin{center}
\begin{tabular}{ c | c | c | c | c} 
 $N={n}_0+{n}_1+{n}_2$ & $q$ & Flat Connection & Gauge Group & IIA closed dual? \\ 
 \hline
 $4+0+0$ & $0$ & $(1,1,1,1)$ & $SU(4)$ & YES\\ 
 $2+1+1$ & $0$ & $(\lambda,\bar\lambda,1,1)$ & $SU(2)\times U(1)^2$ & YES\\  
 $1+3+0$ & $0$ & $(\lambda,\lambda,\lambda,1)$ & $SU(3)\times U(1)$\\   
 $1+0+3$ & $0$ & $(\bar\lambda,\bar\lambda,\bar\lambda,1)$ & $SU(3)\times U(1)$ \\ 
 $0+2+2$ & $0$ & $(\lambda,\lambda,\bar\lambda,\bar\lambda)$ & $SU(2)\times SU(2)\times U(1)$\\ 
 $0+4+0$ & $1$ & $(1,1,1,1)$ & $SU(4)$ \\  
 $1+2+1$ & $1$ & $(\lambda,\bar\lambda,1,1)$ & $SU(2)\times U(1)^2$\\   
 $3+1+0$ & $1$ & $(\bar\lambda,\bar\lambda,\bar\lambda,1)$ & $SU(3)\times U(1)$ & YES\\ 
 $0+1+3$ & $1$ & $(\lambda,\lambda,\lambda,1)$ & $SU(3)\times U(1)$ \\   
 $2+0+2$ & $1$ & $(\lambda,\lambda,\bar\lambda,\bar\lambda)$ & $SU(2)\times SU(2)\times U(1)$ & (YES)\\
 $0+0+4$ & $2$ & $(1,1,1,1)$ & $SU(4)$ \\
 $1+1+2$ & $2$ & $(\lambda,\bar\lambda,1,1)$ & $SU(2)\times U(1)^2$\\   
 $3+0+1$ & $2$ & $(\lambda,\lambda,\lambda,1)$ & $SU(3)\times U(1)$ & YES\\  
 $0+3+1$ & $2$ & $(\bar\lambda,\bar\lambda,\bar\lambda,1)$ & $SU(3)\times U(1)$ \\ 
 $2+2+0$ & $2$ & $(\lambda,\lambda,\bar\lambda,\bar\lambda)$ & $SU(2)\times SU(2)\times U(1)$ & (YES)
 \end{tabular}
\end{center}

We were not able to find smooth, complete $G_2$-manifolds for every flat connection. The reason is that sometimes the solutions to the superpotential conditions have negative values of the Kahler moduli. However, physics suggests that such manifolds ought to exist. We have marked by YES the geometries where a (partial) resolution of type I of section \ref{sec:ConZ3} exists. The cases marked (YES) do admit singular resolutions of type I, though it requires putting flux on vanishing curves, which is not covered by the FHN-theorem as the $CY_{3}$ is singular.\footnote{It should be noted that all the geometries admit resolutions of type II. These however do not have a gauge theory interpretation, and also suffer from the problem of having fluxes on vanishing cycles.}  
\end{example}

\subsection{The $\Z_4$-quotients of the conifold}
For general $p$ and $N$ coprime, the total number of flat connections is
\begin{equation}
{\rm Number\:of\:flat\:connections}=\frac{(N+p-1)!}{N!(p-1)!}={N+p-1\choose N}\:.
\end{equation}
This is also the number of ways to partition $N$ into $p$ distinct integers. However $N$ is now a weighted sum of flux integers for $p>3$ 
\begin{equation}
N={n}_1+{n}_2+..+{n}_0=-\kappa_0+\kappa_1+2\kappa_2+..+2\kappa_{p-2}+\kappa_{p-1}\:.
\end{equation}
We note that in contrast to $p\le3$, the flux $\kappa_0$ associated to the curve $ C_0$ need not be positive anymore.

\begin{example}[$p=4\:,\;N=3$] 
Let us consider the example of the conifold mod $\Z_4$ on the closed string side, with $N=3$ units of flux. In the UV we have four $SU(3)$ gauge theories admitting twenty different flat connections corresponding to flat $U(3)$ connections $\tilde g$ with determinant $\lambda^q$ for $q\in\{0,1,2,3\}$ and $\lambda$ a fourth root of unity. The ones with determinant $\lambda^0=1$ are
\begin{equation}
(1,1,1)\:,\;\;(1,\lambda,\lambda^3)\:,\;\;(1,\lambda^2,\lambda^2)\:,\;\;(\lambda,\lambda,\lambda^2),\;\;(\lambda^3,\lambda^3,\lambda^2)\:,
\end{equation}
with the others given by multiplying these by factors of $\lambda$. We have the following table
\begin{center}
\begin{tabular}{ c | c | c | c | c} 
 $N={n}_0+{n}_1+{n}_2+{n}_3$ & $q$ & Flat $SU(3)$ Connection & Gauge Group & IIA closed dual? \\ 
 \hline
 $3+0+0+0$ & $0$ & $(1,1,1)$ & $SU(3)$ & YES\\ 
 $1+1+0+1$ & $0$ & $(1,\lambda,\lambda^3)$ & $U(1)^2$ \\
 $1+0+2+0$ & $0$ & $(1,\lambda^2,\lambda^2)$ & $SU(2)\times U(1)$ \\ 
 $0+2+1+0$ & $0$ & $(\lambda,\lambda,\lambda^2)$ & $SU(2)\times U(1)$ \\
 $0+0+1+2$ & $0$ & $(\lambda^2,\lambda^3,\lambda^3)$ & $SU(2)\times U(1)$ \\
 $0+0+0+3$ & $1$ & $(1,1,1)$ & $SU(3)$ \\
 $1+0+1+1$ & $1$ & $(1,\lambda,\lambda^3)$ & $U(1)^2$ \\
 $0+2+0+1$ & $1$ & $(1,\lambda^2,\lambda^2)$ & $SU(2)\times U(1)$ \\
 $2+1+0+0$ & $1$ & $(\lambda,\lambda,\lambda^2)$ & $SU(2)\times U(1)$ & (YES)\\
 $0+1+2+0$ & $1$ & $(\lambda^2,\lambda^3,\lambda^3)$ & $SU(2)\times U(1)$ \\
 $0+0+3+0$ & $2$ & $(1,1,1)$ & $SU(3)$ \\
 $0+1+1+1$ & $2$ & $(1,\lambda,\lambda^3)$ & $U(1)^2$ \\
 $2+0+1+0$ & $2$ & $(1,\lambda^2,\lambda^2)$ & $SU(2)\times U(1)$ & YES \\ 
 $1+0+0+2$ & $2$ & $(\lambda,\lambda,\lambda^2)$ & $SU(2)\times U(1)$ \\
 $1+2+0+0$ & $2$ & $(\lambda^2,\lambda^3,\lambda^3)$ & $SU(2)\times U(1)$ \\
 $0+3+0+0$ & $3$ & $(1,1,1)$ & $SU(3)$ \\
 $1+1+1+0$ & $3$ & $(1,\lambda,\lambda^3)$ & $U(1)^2$ \\
 $0+1+0+2$ & $3$ & $(1,\lambda^2,\lambda^2)$ & $SU(2)\times U(1)$ & \\ 
 $0+0+2+1$ & $3$ & $(\lambda,\lambda,\lambda^2)$ & $SU(2)\times U(1)$ \\
 $2+0+0+1$ & $3$ & $(\lambda^2,\lambda^3,\lambda^3)$ & $SU(2)\times U(1)$ & (YES)
 \end{tabular}
\end{center}
\end{example}

\begin{example}[$p=4\:,\;N=5$] 
Having looked at an example with $N<p$, we also an example with $N>p$. We again take $p=4$ but $N=5$. There is now a total of $56$ flat $U(5)$ connections. Here we list the table with the geometries corresponding to the fourteen flat connections of unit determinant.
\begin{center}
\begin{tabular}{ c | c | c | c | c} 
 $N={n}_0+{n}_1+{n}_2+{n}_3$ & $q$ & Flat Connection & Gauge Group & IIA closed dual? \\ 
 \hline
 $5+0+0+0$ & $0$ & $(1,1,1,1,1)$ & $SU(5)$ & YES\\ 
 $1+4+0+0$ & $0$ & $(1,\lambda,\lambda,\lambda,\lambda)$ & $SU(4)\times U(1)$ \\
 $1+0+4+0$ & $0$ & $(1,\lambda^2,\lambda^2,\lambda^2,\lambda^2)$ & $SU(4)\times U(1)$ \\
 $1+0+0+4$ & $0$ & $(1,\lambda^3,\lambda^3,\lambda^3,\lambda^3)$ & $SU(4)\times U(1)$ \\
 $3+1+0+1$ & $0$ & $(1,1,1,\lambda,\lambda^3)$ & $SU(3)\times U(1)^2$ & YES \\
 $0+1+1+3$ & $0$ & $(\lambda,\lambda^2,\lambda^3,\lambda^3,\lambda^3)$ & $SU(3)\times U(1)^2$ \\
 $0+3+1+1$ & $0$ & $(\lambda,\lambda,\lambda,\lambda^2,\lambda^3)$ & $SU(3)\times U(1)^2$ \\
 $3+0+2+0$ & $0$ & $(1,1,1,\lambda^2,\lambda^2)$ & $SU(3)\times SU(2)\times U(1)$ & YES\\
 $0+2+3+0$ & $0$ & $(\lambda,\lambda,\lambda^2,\lambda^2,\lambda^2)$ & $SU(3)\times SU(2)\times U(1)$ \\
 $0+0+3+2$ & $0$ & $(\lambda^2,\lambda^2,\lambda^2,\lambda^3,\lambda^3)$ & $SU(3)\times SU(2)\times U(1)$ \\
 $2+2+1+0$ & $0$ & $(1,1,\lambda,\lambda,\lambda^2)$ & $SU(2)^2\times U(1)^2$ \\
 $1+2+0+2$ & $0$ & $(1,\lambda,\lambda,\lambda^3,\lambda^3)$ & $SU(2)^2\times U(1)^2$ \\
 $2+0+1+2$ & $0$ & $(1,1,\lambda^2,\lambda^3,\lambda^3)$ & $SU(2)^2\times U(1)^2$ \\
 $1+1+2+1$ & $0$ & $(1,\lambda,\lambda^2,\lambda^2,\lambda^3)$ & $SU(2)\times U(1)^3$ 
 \end{tabular}
\end{center}
Additionally there are five more flat $U(5)$ connections with non-unit determinant which admit (partial) dual resolutions. These correspond to $SU(5)$ gauge theories where $q\neq0$. We list them here,
\begin{center}
\begin{tabular}{ c | c | c | c | c} 
 $N={n}_0+{n}_1+{n}_2+{n}_3$ & $q$ & Flat Connection & Gauge Group & IIA closed dual? \\ 
 \hline
 $4+1+0+0$ & $1$ & $(1,\lambda^3,\lambda^3,\lambda^3,\lambda^3)$ & $SU(4)\times U(1)$ & YES\\ 
 $4+0+1+0$ & $2$ & $(1,\lambda^2,\lambda^2,\lambda^2,\lambda^2)$ & $SU(4)\times U(1)$ & YES\\ 
 $4+0+0+1$ & $3$ & $(1,\lambda,\lambda,\lambda,\lambda)$ & $SU(4)\times U(1)$ & YES\\  
 $3+0+1+1$ & $1$ & $(\lambda,\lambda^2,\lambda^3,\lambda^3,\lambda^3)$ & $SU(3)\times U(1)^2$ & (YES)\\  
 $3+1+1+0$ & $3$ & $(\lambda,\lambda,\lambda,\lambda^2,\lambda^3)$ & $SU(3)\times U(1)^2$ & (YES)\\  
 \end{tabular}
\end{center}

\end{example}

\subsection{Further Physical checks} 

Since the UV gauge group contains some number of essentially free $U(1)$ factors, this gauge symmetry must exist at each point in the corresponding component of the moduli space. Can we see this gauge symmetry on the IR side? In Type IIA theory the $U(1)$ gauge symmetry can arise from the 3-form RR gauge potential $C_3$ via a Kaluza-Klein ansatz that involves a harmonic 2-form on the Calabi-Yau threefold. In fact, in order to be a genuine gauge symmetry, the harmonic 2-form must be $L^2$-normalisable. Remarkably, there is a perfect agreement for all of the cases marked YES. This is because the $L^2$-normalisable harmonic 2-forms are essentially in correspondence with compact divisors in the Calabi-Yau threefold. This also matches the number of $L^{2}$-normalizable harmonic two-forms in the corresponding ALC $G_2$ manifolds, a result which also follow from the more general analysis of \cite{hausel2004hodge}. This provides further evidence that the $M$-theory flop transitions might occur on a smooth quantum moduli space.

\subsection{The classical moduli space}
We considered IIA theory on a $\Z_p$ quotient $T^\ast (S^3/\Z_p)$ of the deformed conifold with N D6 branes wrapped around $S^3/\Z_p$ and a RR $1$-form gauge potential determined by $q\in \{0,1,\dots, p-1\}$.

The IIA configuration is up-lifted to $M$-theory on the $G_2$ orbifold $(S^3\times \R^4)/\Gamma_{p,N,q}$ asymptotic to a cone $C(S^3\times S^3)/\Gamma_{p,N,q}$. In order to understand the classical moduli space of this theory we need to understand moduli spaces of $G_2$ metrics asymptotic to the same cone.

Note that even if cones associated with different choices of $p,N,q$ can be isometric, they are always distinct $G_2$ cones. For example, if $p=2$ and $N$ is odd then $\Gamma_{2,N,1}$ can be identified with $\Gamma_{2,N,0}$ by exchanging the two factors of $S^3\times S^3$; while this transformation is an isometry of the Bryant-Salamon cone $C(S^3\times S^3)$, it is not a $G_2$ automorphism. Thus triples $(p,N,q)$ label different classical moduli spaces.

Orbifolds of the standard Bryant-Salamon smooth $G_2$ structures on $S^3\times \R^4$ give rise to three 1-dimensional branches of the classical moduli spaces. If $q=0$ these three branches are
\begin{enumerate}
\item  $S^3/\Z_p \times \R^4/\Z_N$, corresponding to the $M$-theory lift of a IIA configuration consisting of the $\Z_p$ quotient of the deformed conifold with $N$ D6 branes wrapped on $S^3/\Z_p$;
\item $S^3/\Z_N \times \R^4/\Z_p$, corresponding to the $M$-theory lift of a IIA configuration consisting of the $\Z_p$ quotient of the resolved conifold with N units of RR flux through the $\mathbb{P}^1$;
\item the smooth $G_2$ manifold $(S^3 \times \R^4) / (\Z_p\times\Z_N)$ \cite{Friedmann:2002ct}.
\end{enumerate} 
The codimension-4 ADE singularities in cases (i) and (ii) give rise to different backgrounds labeled by $SU(N)$ or $SU(2)$ Wilson lines on $S^3/\Z_p$ and $S^3/\Z_N$ respectively.      

The FHN theorem \cite{foscolo2017complete} allows us to construct additional branches of the classical moduli space arising from $M$-theory uplifts of IIA configurations consisting of a resolution of the hyperconifold $\mathcal{C}/\Z_p$ with appropriate configurations of RR fluxes. As we have seen, only special choices of flux configurations allow the existence of a K\"ahler form $\omega$ such that $c_1 \wedge \omega =0$. Moreover, for any such allowable choice flux choice there is a unique $\omega$ up to scale such that $c_1 \wedge \omega =0$. Thus \cite{foscolo2017complete} gives $G_2$-metrics that depend on a single parameter up to scale: if we rescale the metrics to fix the size of the $M$-theory circle at infinity, the free parameter is identified with the single allowable K\"ahler parameter $\omega$.

Strictly speaking the construction of \cite{foscolo2017complete} guarantees the existence of $G_2$ metrics only when the $M$-theory circle is much smaller than the K\"ahler parameters. Thus these $G_2$ metrics are asymptotic to the cone $C(S^3\times S^3)/\Gamma_{p,N,q}$ only topologically and not at the metric level. However, precisely because these $G_2$ metrics move in $1$-parameter families up to scale, it is not unreasonable to assume the existence of $G_2$ metrics for any size of the $M$-theory circle: in the limit of large radius we obtain a $G_2$ metric (unique up to scale) asymptotic to the $G_2$ cone $C(S^3\times S^3)/\Gamma_{p,N,q}$. In the simplest case $p=2$ the existence of these asymptotically conical $G_2$ metrics was established in \cite{foscolo2018infinitely}. 

We conclude that every time we have a YES in the tables of this section, we add a $1$-dimensional branch to the classical moduli space. The number of independent $L^2$-integrable harmonic $2$-forms on each of these branches is exactly $p-1$, matching the number of free $U(1)$ factors on the UV side.

The configuration of fluxes labelled with (YES) imply that the only solution of $c_1\wedge\omega=0$ is a form $\omega$ at the boundary of the K\"ahler cone. While we cannot apply the construction of \cite{foscolo2017complete} directly, we expect these configurations to give rise to additional $1$-dimensional branches of the classical moduli space corresponding to singular $G_2$ spaces asymptotic to $C(S^3\times S^3)/\Gamma_{p,N,q}$.

An open question is whether the classical moduli space contains additional branches. By a result of Karigiannis-Lotay \cite{Karigiannis2012Deformation} every smooth $G_2$ manifold asymptotic to $C(S^3\times S^3)$ or one of its quotients inherits all the $G_2$ symmetries of the cone. When $p=2$ the symmetries of $C(S^3\times S^3)/\Gamma_{2,N,q}$ contain $SU(2)\times SU(2) \times U(1)$: the classification of complete $G_2$ metrics with $SU(2)\times SU(2) \times U(1)$ symmetry in \cite{foscolo2018infinitely} guarantees that the classical moduli space contains no other branches than the ones listed above. For $p>2$, the symmetry group of $C(S^3\times S^3)/\Gamma_{p,N,q}$ is only $SU(2)\times U(1)\times U(1)$ and no classification of $G_2$ metrics with this symmetry is currently available. We note that since the symmetry group contains $T^3$, classifying $G_2$ geometries asymptotic to $C(S^3\times S^3)/\Gamma_{p,N,q}$ is equivalent to classifing $(p,q,r)$ webs with four fixed external legs (determined by $p,N,q$).

\begin{figure}[H]
\begin{center}
\subfloat{\includegraphics[width = 4.5in]{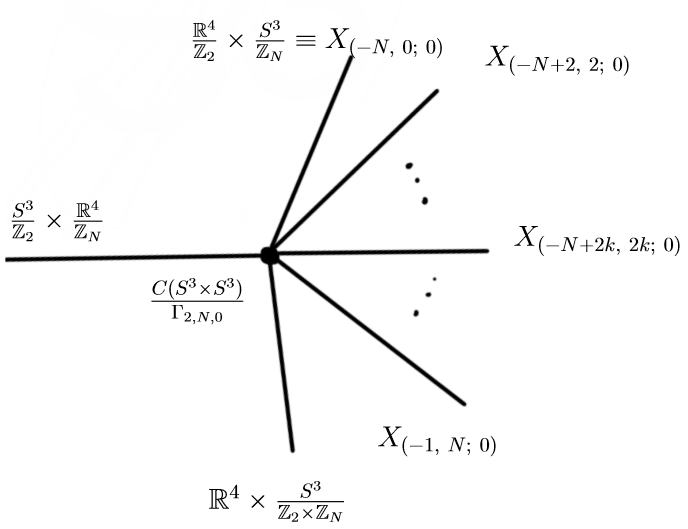}}
\caption{The classical moduli space for $p=2$, $N$ odd, and $q=0$.}
\label{clasmodulispace}
\end{center}
\end{figure}

\section{Outlook}

There are some potentially interesting avenues for further study which arise. One direction would be to investigate non-Abelian $\Gamma_{ADE}$ quotients. The new $G_2$-manifolds constructed this way might have additional features arising from the $SO(2N)$ or $E_6, E_7$ or $E_8$ gauge symmetries and corresponding flat connections, although one loses the use of toric geometry to solve the superpotential conditions.

Much more generally, the FHN theorem applies to {\it any} complete asymptotically conical Calabi-Yau threefold for which the superpotential conditions can be satisfied. Hence, there will likely be many more applications of it in the context of $M$-theory/Type IIA duality.

\bigskip
\large
\noindent
{\bf {\sf Acknowledgements.}}
\normalsize

We would like to thank Mark Haskins and Michele del Zotto for comments and suggestions during the course of this work. LF would like to thank Philipp Breen-Ulbrich for help with some of the computations for the tables in Section 4. EES thanks the International Centre for Theoretical Physics Trieste for their hospitality during the bulk of the project. The work of BSA, MN and ESS is supported by a grant from the Simons Foundation (\#488569, Bobby Acharya). LF is supported by a Royal Society University Research Fellowship.

\bigskip

\bibliographystyle{ieeetr}


\end{document}